\documentclass[12pt]{iopart}

\usepackage{graphicx}

\begin{document}

\title[Nonlinear vibration transfer in torsion pendulums]{Nonlinear vibration transfer in torsion pendulums}

\author{Tomofumi Shimoda \& Masaki Ando}

\address{Department of physics, University of Tokyo, 7-3-1, Hongo, Bunkyo-ku, Tokyo 113-0033, Japan}
\ead{shimoda@granite.phys.s.u-tokyo.ac.jp}
\vspace{10pt}
\begin{indented}
\item[]December 2018
\end{indented}

\begin{abstract}
Torsion pendulums have been widely used in physical experiments, because their small restoring forces are suitable for tiny force measurement.
Recently, some applications such as low-frequency gravity gradiometers have been proposed by focusing on their low resonant frequencies.
Torsion pendulums with low resonant frequencies enable the suspended masses to be isolated from the ground, allowing for good response to the fluctuation of local gravity field at low frequencies.
However, translational ground vibration can be transferred to the horizontal rotation of the torsion pendulum nonlinearly.
This effect can have a non-negligible contribution to the sensitivity of torsion pendulums as gravity gradiometers.
This paper evaluates the amount of nonlinear vibration noise, and discusses how to reduce it. 
\end{abstract}

%
\vspace{2pc}

\section{Introduction}
A torsion pendulum is a long-standing tool in physical experiments. 
Its small restoring force enables tiny force measurements such as those required in gravitational experiments.
Hence it has been used for the test of gravitational inverse-square law \cite{ISL_Washington,ISL_WHT}, measurement of Newton's gravitational constant $G$ (e.g. \cite{G_Quinn, G_LCT, G_cryo}) and other experiments.

\begin{figure}
\begin{center}
	\begin{tabular}{ll}
	\begin{minipage}{0.45\hsize}
	\begin{center}
	\includegraphics[width=7cm]{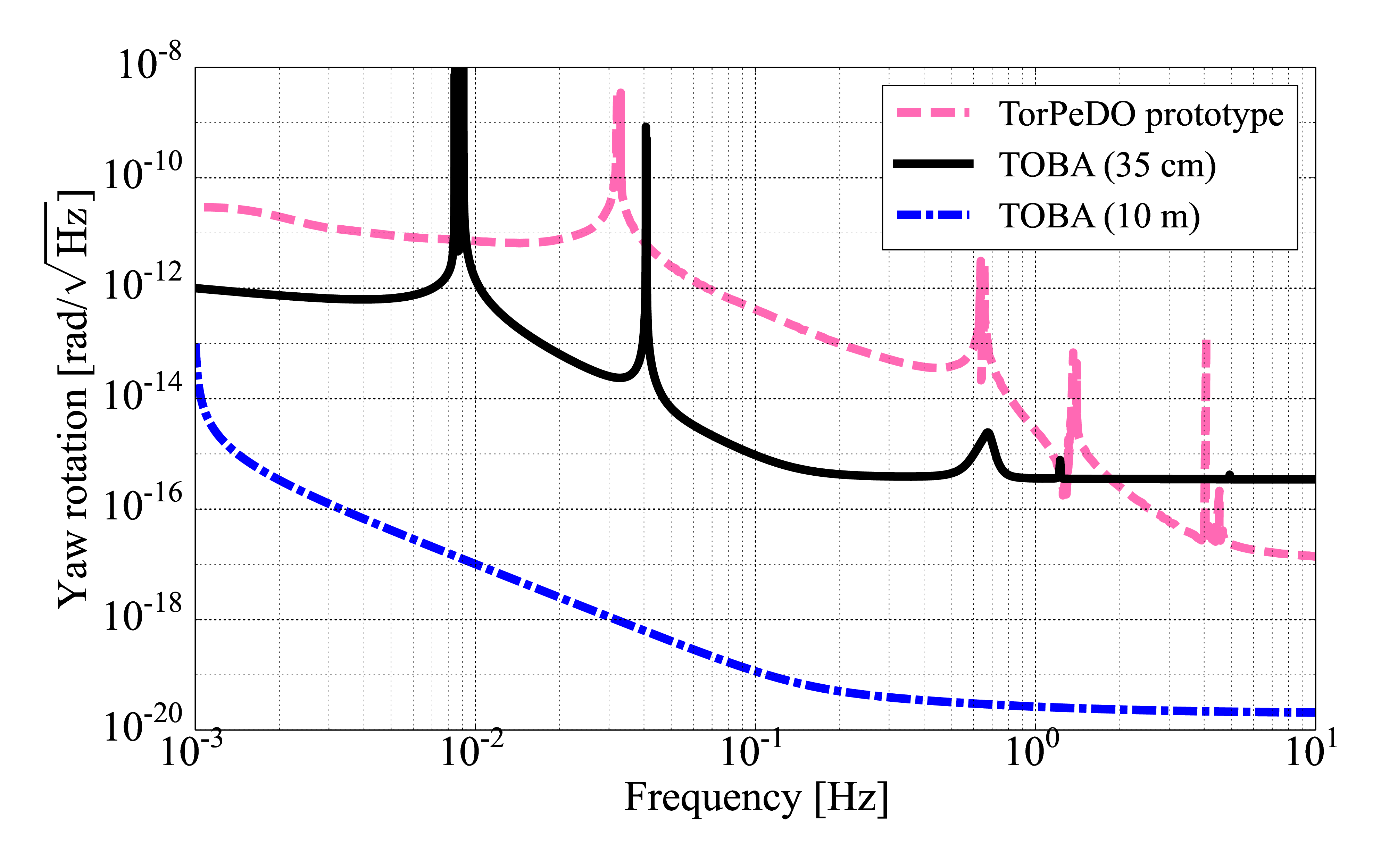}
	(a)
	\end{center}
	\end{minipage}
	\begin{minipage}{0.45\hsize}
	\begin{center}
	\includegraphics[width=7cm]{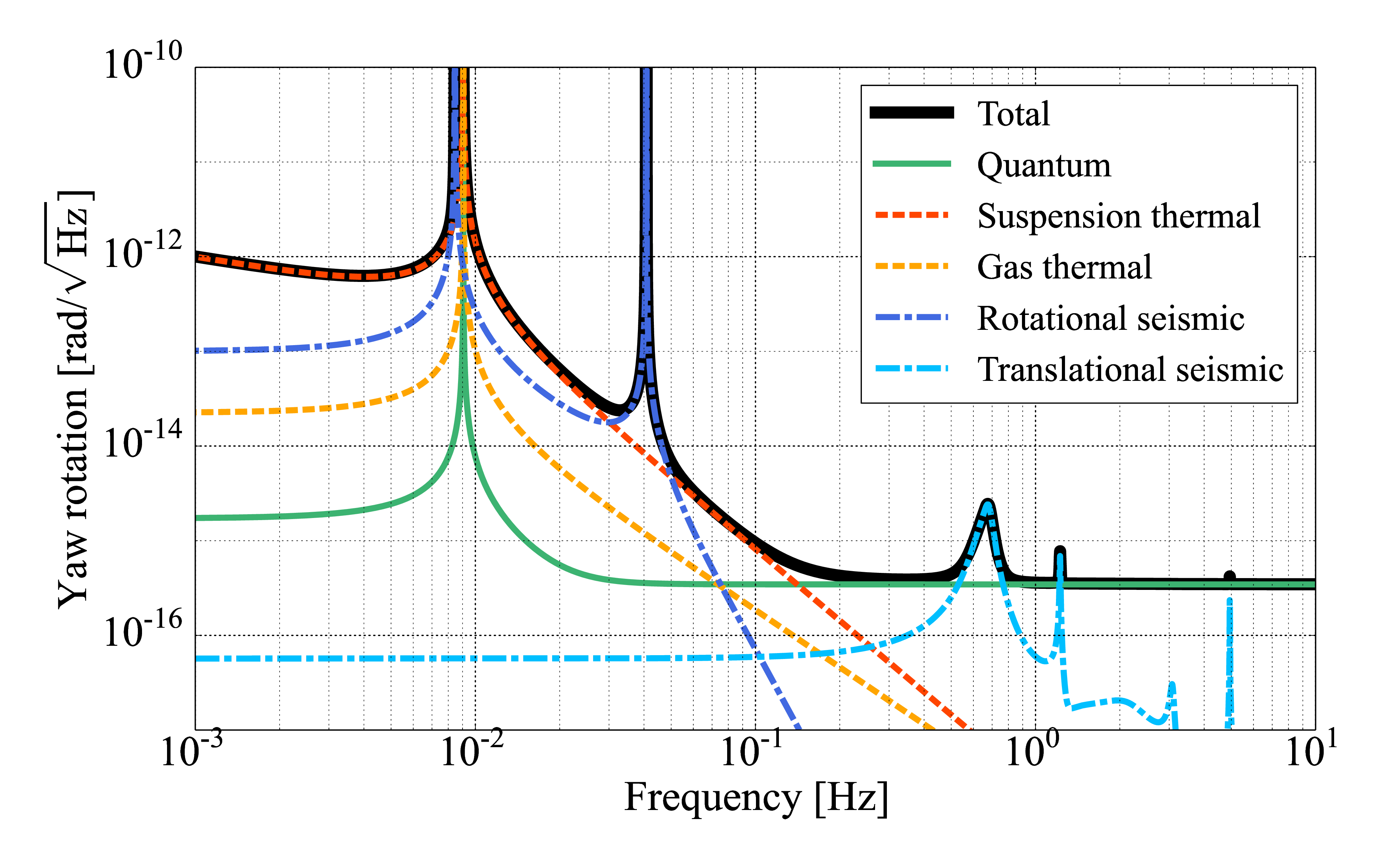}
	(b)
	\end{center}
	\end{minipage}
	\end{tabular}
\caption{\label{fig:target} (a) The target sensitivities of proposed gravity gradiometers and/or their prototypes : prototype of TorPeDO (dashed pink), TOBA with 35 cm bars (solid black), and TOBA with 10 m bars (dot-dashed blue).  (b) The noise budget of TOBA with 35 cm bars. It is mainly limited by quantum noise (solid green) and thermal noise of the suspension wires (dashed red), and partly limited by rotational (dot-dashed blue) and translational (dot-dashed light blue) seismic noise, which are linearly transferred to the horizontal rotation of the bars. Thermal noise of residual gas (dashed orange) is not dominant.}
\end{center}
\end{figure}

Recently, Torsion-Bar Antenna (TOBA), a local gravity gradiometer using a torsion pendulum, has also been proposed and is being developed \cite{TOBA, Phase1TOBA, Phase2TOBA}.
It utilizes the low resonant frequency of a torsion pendulum ($\sim$ a few mHz), which leads the pendulum to behave like a free-falling body down to the resonant frequency.
This enables a good response to gravitational waves (GWs) and passive isolation of rotational vibration of the suspension point above the resonant frequency.
Hence a torsion pendulum can have good sensitivity to GWs at low frequencies even on land.
This configuration is also expected to be useful for earthquake early warning (EEW) by measuring gravity perturbations \cite{EEW, EEW_JPM2016, EEW_MV2017}.
The similar detector TorPeDO (Torsion Pendulum Dual Oscillator) is currently under development for the purpose of such terrestrial gravity measurements \cite{TorPeDO,TorPeDO_EEW}.
Fig. \ref{fig:target} shows the target sensitivities of these detectors and/or their prototypes \cite{TOBA, TorPeDO}.
The target noise level is roughly about $10^{-15}$ rad/$\sqrt{\rm Hz}$ around 0.1 Hz for EEW, and $10^{-19}$ rad/$\sqrt{\rm Hz}$ for GWs.
In the Fig. \ref{fig:target} (b), the main noise sources of TOBA with 35 cm bars are plotted for comparison against the nonlinear vibration noise discussed in this paper.

In these applications, seismic noise is one of the noise sources which can limit sensitivity.
The rotational motion of the ground can be easily isolated with a multi-stage torsion pendulum, which works as a passive vibration isolation system.
The horizontal rotational of the pendulum induced by the translational motion of the ground, called ``cross-coupling'', has also been investigated.
It was found that the linear cross-coupling transfer is caused by the asymmetry of the system, so it can be suppressed by improving the symmetry \cite{TOBA_SCC}.
However, even if the pendulum is completely symmetric in its stationary state, vibration of the ground induces momentary asymmetry of the pendulum, which creates nonlinear cross-coupling.
This effect has not been studied so far in terms of noise of the pendulum.

In this paper, the nonlinear vibration transfer in a torsion pendulum is investigated.
The principle of this transfer is explained in Sec. \ref{sec:principle}, which is evaluated for some cases, followed by a discussion on how to reduce it in Sec. \ref{sec:reduction} and \ref{sec:discussion}.

\section{Nonlinear vibration transfer}\label{sec:principle}
\begin{figure}
\begin{center}
\includegraphics[width=7cm]{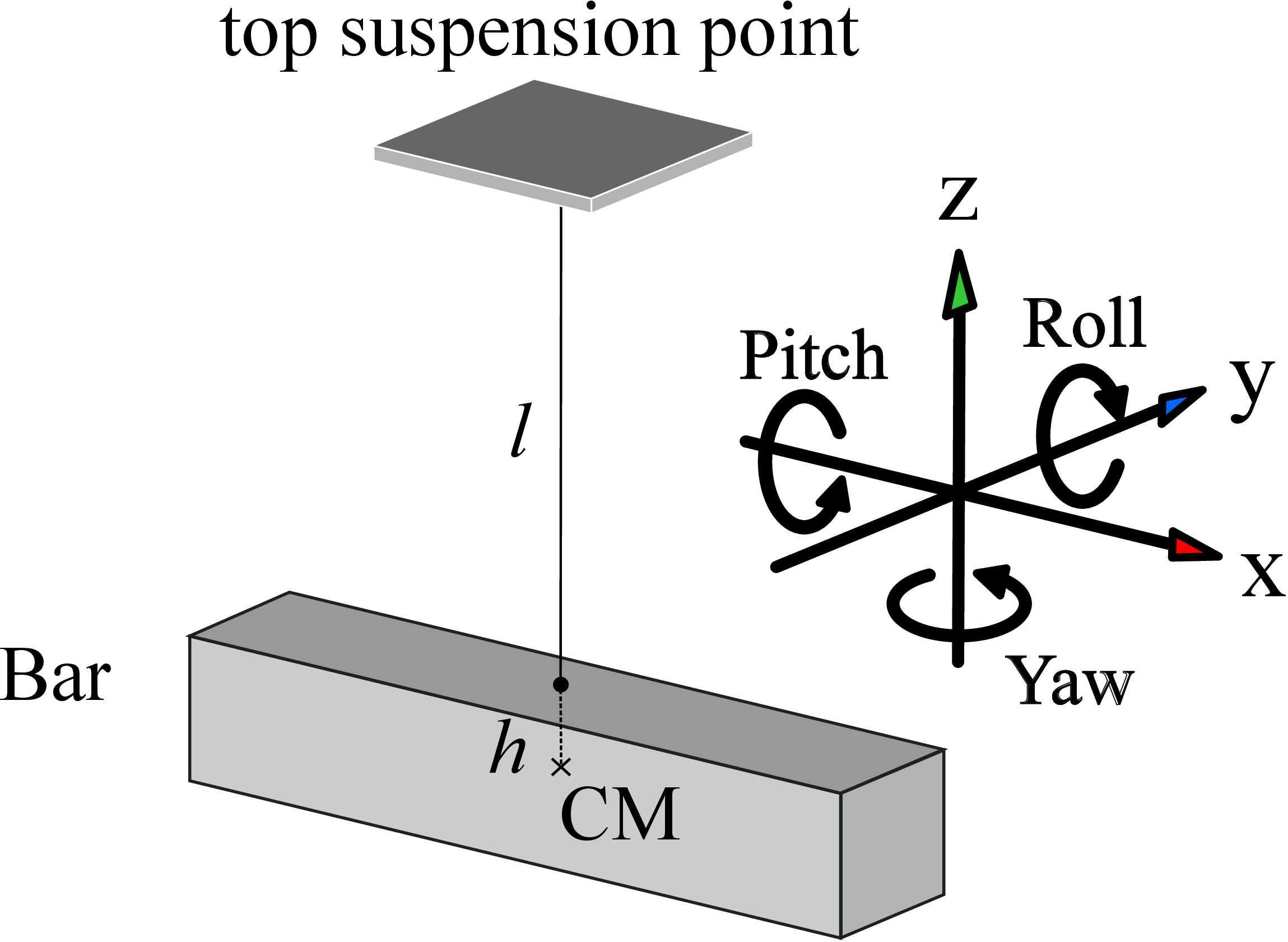}
\caption{\label{fig:model}The model of a torsion pendulum and definition of coordinates.}
\end{center}
\end{figure}

The purpose of this section is to derive the explicit formula of horizontal (Yaw) rotation induced by nonlinear vibration transfer, and see what parameters are important for it.
The definition of the coordinates and parameters of the system are shown in Fig. \ref{fig:model}.
The length of the suspension wire is defined as $l$, and the distance between the suspension point and the center of mass (CM) is $h$.
In the following discussion, $(x, y, z)$ is the position of the CM, and $(\theta_{\rm P}, \theta_{\rm R}, \theta_{\rm Y})$ is the rotational angle around the CM.
The subscripts P, R and Y indicates Pitch, Roll and Yaw, respectively.
The horizontal translations of the ground are expressed as $x_{\rm g}$ and $y_{\rm g}$.

\subsection{Derivation}\label{sec:derivation}
The Lagrangian of the system is
\begin{equation}
\mathcal{L} = \frac{1}{2} m \left( \dot{x}^2 + \dot{y}^2 \right) + \frac{1}{2} \vec{\omega}\cdot\bi{I} \vec{\omega} - mgz
 - \frac{1}{2}\kappa_{\rm Y} \theta_{\rm Y}^2 \label{eq:L}
\end{equation}
and the dissipation function is defined as 
\begin{equation}
\mathcal{D} = \frac{1}{2}\Gamma_{\rm x} \dot{x}^2 + \frac{1}{2}\Gamma_{\rm y} \dot{y}^2 + \frac{1}{2}\Gamma_{\rm P} \dot{\theta}_{\rm P}{}^2 + \frac{1}{2}\Gamma_{\rm R} \dot{\theta}_{\rm R}{}^2 + \frac{1}{2}\Gamma_{\rm Y} \dot{\theta}_{\rm Y}{}^2
\end{equation}
Here we take the kinetic energy, gravitational potential and elastic energy of the torsional mode of the wire into account.
The elastic energy about the bending of the wire is ignored here by assuming that the wire is thin enough. 
$m$ and $\bi{I}$ are the mass and the moment tensor of the bar, 
$\vec{\omega} = \frac{d}{dt} (\theta_{\rm P}, \theta_{\rm R}, \theta_{\rm Y})$ is the angular velocity vector, 
and $\kappa_{\rm Y}$ is the torsional spring constant of the wire. 
$\Gamma_{\alpha}$ ($\alpha={\rm x,y,P,R,Y}$) is a damping coefficient of each degree of freedom.
In the following discussion, the displacement is assumed to be small enough so that the small angle approximation ($\sin\theta\simeq\theta$ and $\cos\theta\simeq1-\theta^2/2$) is valid.  
In Eq. (\ref{eq:L}) and the following equations, $(t)$ of the time-domain variables, such as $x$, $y$, $z$, $\theta_{\rm P}$, $\theta_{\rm R}$ or $\theta_{\rm Y}$, are omitted. 
Since the purpose of this calculation is the nonlinear effect, a time-varying moment tensor $\bi{I}(t)$ has to be used.
Its non-diagonal elements are caused by the Pitch and the Roll rotations induced by the seismic vibration, and is
\begin{equation}
\bi{I}(t) = \left(
	\begin{array}{ccc}
		I_{\rm P} 							& 0								 	& \theta_{\rm R} ( I_{\rm Y} - I_{\rm P} )  \\
		0									& I_{\rm R} 							& \theta_{\rm P} ( I_{\rm R} - I_{\rm Y} ) \\
		\theta_{\rm R} ( I_{\rm Y} - I_{\rm P} ) 	& \theta_{\rm P} ( I_{\rm R} - I_{\rm Y} ) 	& I_{\rm Y}  \\
	\end{array}
	\right) . 
\label{eq:I}
\end{equation}
Here $I_{\rm P}$, $I_{\rm R}$ and $I_{\rm Y}$ are moments of inertia around principal axes of the suspended bar.
The vertical position of the center of mass (CM), $z$, can be calculated geometrically.
As seismically excited vibrations of a typical pendulum considered here are almost always below $10^{-5}$ m/$\sqrt{\rm Hz}$ for translations and $10^{-4}$ rad/$\sqrt{\rm Hz}$ for vertical rotations, as shown later in Fig. \ref{fig:seismic} (b), the lowest order terms of $x/h$, $y/h$, $\theta_{\rm P}$ and $\theta_{\rm R}$ are larger than the higher order terms by at least two orders of magnitude in the following calculations.
By ignoring the third and higher order terms, the time-domain expression of $z$ is 
\begin{equation}
\fl z \simeq \frac{1}{2l} \left\{ \left( x + h \theta_{\rm R} + h \theta_{\rm P} \theta_{\rm Y} - x_{\rm g} \right)^2 + \left( y - h \theta_{\rm P} + h \theta_{\rm R} \theta_{\rm Y} - y_{\rm g} \right)^2  \right\} + \frac{1}{2} h \left( \theta_{\rm P}{}^2 + \theta_{\rm R}{}^2 \right) .
\label{eq:z}
\end{equation}
The non-diagonal elements of Eq. (\ref{eq:I}) and the cross terms between $\theta_{\rm Y}$ and $\theta_{\rm P}$ or $\theta_{\rm R}$ in Eq. (\ref{eq:z}) are the source of nonlinear vibration transfer to Yaw rotation.

The Euler-Lagrange equations can be derived from (\ref{eq:L}) with (\ref{eq:I}) and (\ref{eq:z}).
The equations for $x$ and $y$ are
\begin{equation}
m\ddot{x} = -\frac{mg}{l} \left( x + h \theta_{\rm R} + h \theta_{\rm P} \theta_{\rm Y} - x_{\rm g} \right) - \Gamma_{\rm x} \dot{x} \label{eq:x} 
\end{equation}
and
\begin{equation}
m\ddot{y} = -\frac{mg}{l} \left( y - h \theta_{\rm P} + h \theta_{\rm R} \theta_{\rm Y} - y_{\rm g} \right) - \Gamma_{\rm y} \dot{y}. \label{eq:y} 
\end{equation}
By using these equations, we get the equation for $\theta_{\rm Y}$ as
\begin{equation}
\fl
\kappa_{\rm Y} \theta_{\rm Y} + I_{\rm Y} \ddot{\theta}_{\rm Y} 
=  
- ( I_{\rm R} - I_{\rm Y} ) \theta_{\rm P} \ddot{\theta_{\rm R}} - ( I_{\rm Y} - I_{\rm P} ) \ddot{\theta_{\rm P}} \theta_{\rm R} 
+ ( I_{\rm P} - I_{\rm R} ) \dot{\theta}_{\rm P} \dot{\theta_{\rm R}}
- mh ( \ddot{x} \theta_{\rm P} + \ddot{y} \theta_{\rm R} ).
\end{equation}
Here the dissipation terms are assumed to be much smaller than the other terms, ${\it e.g.}$ $m\ddot{x} \gg \Gamma_{\rm x} \dot{x}$, and ignored for simplicity.
In the frequency-domain, the Fourier-transformed Yaw angle $\tilde{\theta}_{\rm Y}$ is
\begin{eqnarray}
\fl
\tilde{\theta}_{\rm Y} 
= 
\frac{1}{\kappa_{\rm Y} - I_{\rm Y} \omega^2}
\left[
( I_{\rm R} - I_{\rm Y} ) \tilde{\theta}_{\rm P} \ast \left( \omega^2 \tilde{\theta}_{\rm R} \right) +
( I_{\rm Y} - I_{\rm P} ) \left( \omega^2 \tilde{\theta}_{\rm P} \right) \ast \tilde{\theta}_{\rm R} \right. \nonumber \\
\left.
- ( I_{\rm P} - I_{\rm R} ) \left( \omega \tilde{\theta}_{\rm P} \right) \ast \left( \omega \tilde{\theta}_{\rm R} \right)
+ mh \left\{ \left( \omega^2 \tilde{x} \right) \ast \tilde{\theta}_{\rm P} + \left( \omega^2 \tilde{y} \right) \ast \tilde{\theta}_{\rm R} \right\}
\right].
\label{eq:theta_Y}
\end{eqnarray}
Here ``$\ast$'' means frequency convolution defined by
\begin{equation}
(F\ast G)(f) = \int_{-\infty}^{\infty} F(x) G(f-x) dx \hspace{20pt} \left( F(f), G(f) : {\rm functions} \right).
\end{equation}
The noise spectrum of Yaw rotation, which is the main target of this paper, can be calculated from $\tilde{x}$, $\tilde{y}$, $\tilde{\theta}_{\rm P}$ and $\tilde{\theta}_{\rm R}$ by using Eq. (\ref{eq:theta_Y}).
The terms inside the square brackets are the nonlinear torque noise.
In calculating $\tilde{x}$, $\tilde{y}$, $\tilde{\theta}_{\rm P}$ and $\tilde{\theta}_{\rm R}$, the nonlinear effect does not have to be considered.
As shown in Eq. (\ref{eq:x}) and (\ref{eq:y}), the nonlinear terms (the third term in the bracket of each equation) are smaller than the other terms by the order of $\theta_{\rm Y}$, which is at most $10^{-5}$ rad rms as calculated later (Fig. \ref{fig:calculation}).
Hence they are dominated by linearly induced motions.
Under this condition, the equations of motion about $x$, $y$, $\theta_{\rm P}$ and $\theta_{\rm R}$ are 
\begin{eqnarray}
m\ddot{x} = -\frac{mg}{l} \left( x + h \theta_{\rm R} - x_{\rm g} \right) - \Gamma_{\rm x} \dot{x},  \label{eq:x_lin} \\
m\ddot{y} = -\frac{mg}{l} \left( y - h \theta_{\rm P} - y_{\rm g} \right) - \Gamma_{\rm y} \dot{y},  \label{eq:y_lin} \\ 
I_{\rm P} \ddot{\theta}_{\rm P} = \frac{mgh}{l} \left( y - h \theta_{\rm P} - y_{\rm g} \right) - mgh\theta_{\rm P} - \Gamma_{\rm P} \dot{\theta}_{\rm P}, \label{eq:P_lin} \\
I_{\rm R} \ddot{\theta}_{\rm R} = -\frac{mgh}{l} \left( x + h \theta_{\rm R} - x_{\rm g} \right) - mgh\theta_{\rm R} - \Gamma_{\rm R} \dot{\theta}_{\rm R}. \label{eq:R_lin} 
\end{eqnarray}
Assuming $l\gg h$, $\tilde{x}$, $\tilde{y}$, $\tilde{\theta}_{\rm P}$ and $\tilde{\theta}_{\rm R}$ are approximated to 
\begin{eqnarray}
\tilde{x} \simeq \frac{f_x{}^2}{f_x{}^2 + i \frac{f_x}{Q_x}f - f^2} \tilde{x}_{\rm g}, \label{eq:x_fourier} \\
\tilde{y} \simeq \frac{f_y{}^2}{f_y{}^2 + i \frac{f_y}{Q_y}f - f^2} \tilde{y}_{\rm g}, \label{eq:t_fourier} \\
\tilde{\theta}_{\rm P} \simeq \frac{4\pi^2f_y{}^2 f_{\rm P}{}^2 f^2 }{g \left(f_y{}^2 + i \frac{f_y}{Q_y}f - f^2 \right) \left(f_{\rm P}{}^2 + i \frac{f_{\rm P}}{Q_{\rm P}}f - f^2 \right)} \tilde{y}_{\rm g}, \label{eq:P_fourier}\\
\tilde{\theta}_{\rm R} \simeq - \frac{4\pi^2f_x{}^2 f_{\rm R}{}^2 f^2 }{g \left(f_x{}^2 + i \frac{f_x}{Q_x}f - f^2 \right) \left(f_{\rm R}{}^2 + i \frac{f_{\rm R}}{Q_{\rm R}}f - f^2 \right)} \tilde{x}_{\rm g} .\label{eq:R_fourier}
\end{eqnarray}
Here the damping coefficients $\Gamma_{\alpha}$ ($\alpha={\rm x,y,P,R}$) have been converted into the Q factors $Q_{\alpha}$.
The resonant frequencies $f_x$, $f_y$, $f_{\rm P}$ and $f_{\rm R}$ are 
\begin{equation}
f_x \simeq f_y \simeq \frac{1}{2\pi}\sqrt{\frac{g}{l+h}}, \hspace{20pt} f_{\rm P} \simeq \frac{1}{2\pi}\sqrt{\frac{mgh}{I_{\rm P}}}, \hspace{10pt}{\rm and}\hspace{10pt} f_{\rm R} \simeq \frac{1}{2\pi}\sqrt{\frac{mgh}{I_{\rm R}}}. \label{eq:resonances}
\end{equation}

As Eq. (\ref{eq:theta_Y}) shows, the nonlinear torque depends on the mass, the moments of inertia, the height of suspension point, as well as the amplitudes of $x$, $y$ translations and Roll, Pitch rotations.



\subsection{Calculation}\label{sec:calculation}
\begin{figure}
\begin{center}
	\begin{tabular}{ll}
	\begin{minipage}{0.45\hsize}
	\begin{center}
	\includegraphics[width=7cm]{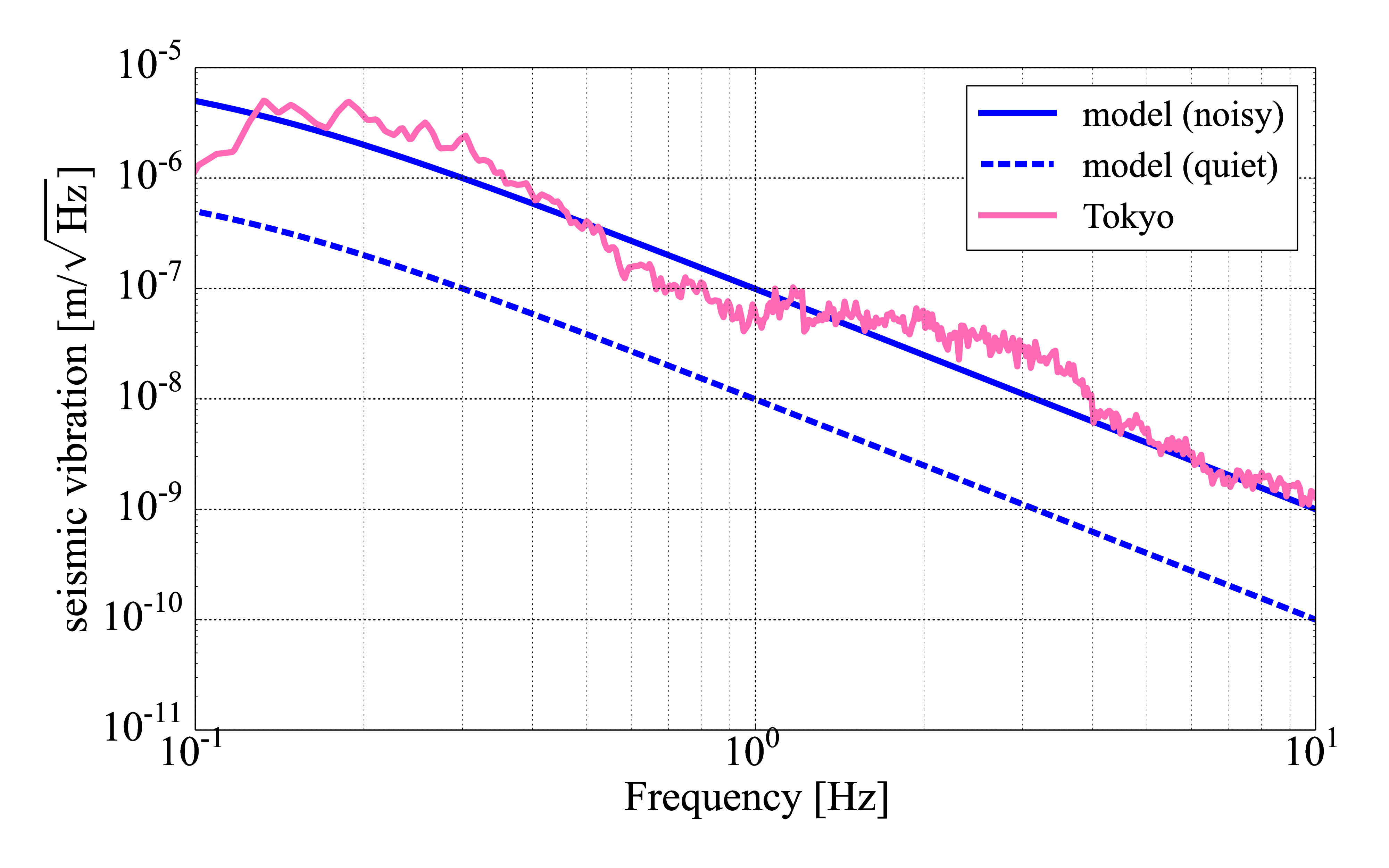}
	(a)
	\end{center}
	\end{minipage}
	\begin{minipage}{0.45\hsize}
	\begin{center}
	\includegraphics[width=7cm]{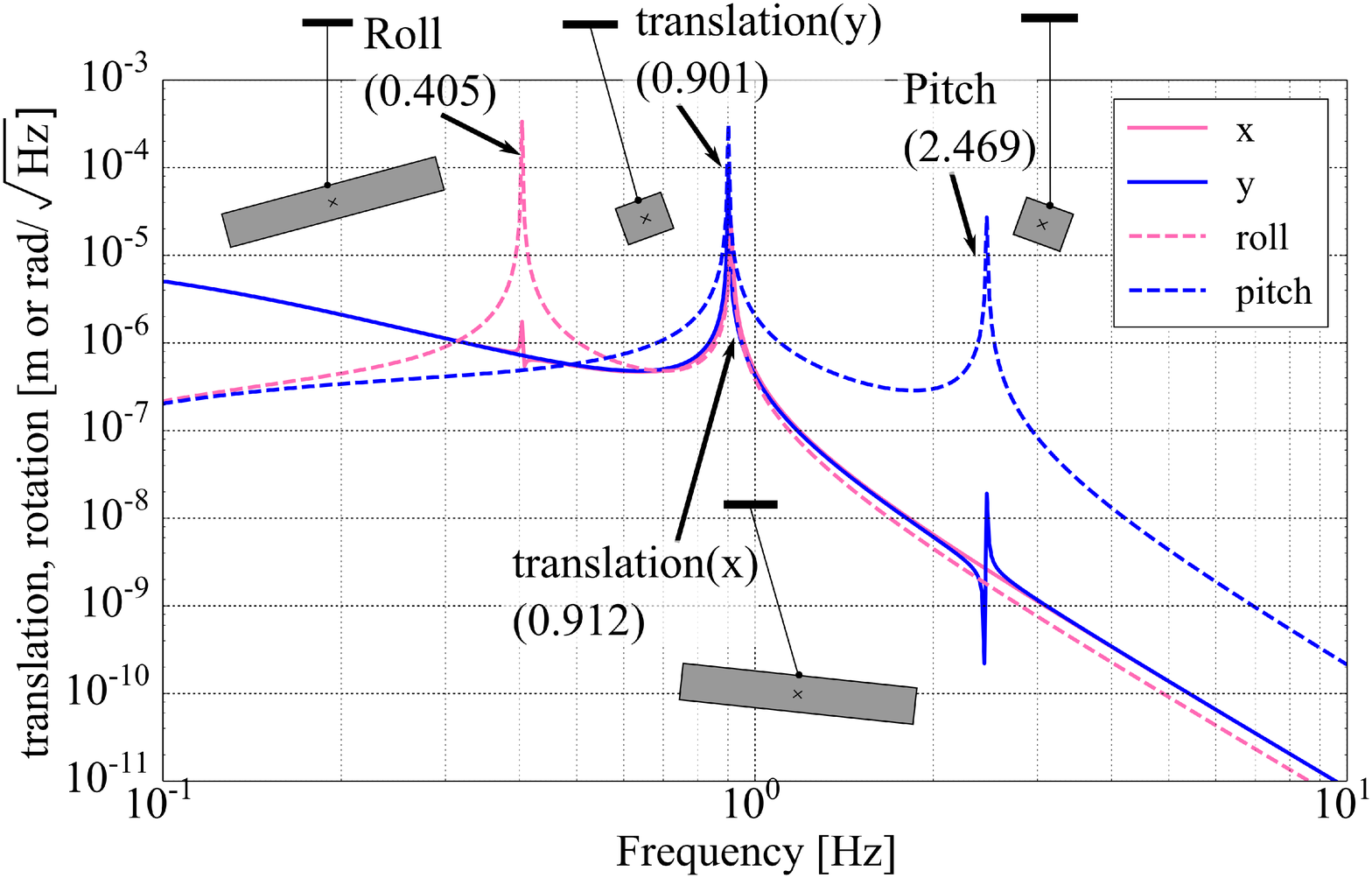}
	(b)
	\end{center}
	\end{minipage}
	\end{tabular}
\caption{\label{fig:seismic} (a) The amplitude spectral density (ASD) of assumed seismic vibration models for a noisy case (blue solid line) and a quiet case (blue dashed line). Vibrations measured in Tokyo (pink) is also shown for comparison. (b) The respective pendulum's translations ($x$ and $y$) and rotations (Roll and Pitch) for noisy model of seismic vibrations. Four resonant modes are drawn and their frequencies are indicated in the bracket.}
\end{center}
\end{figure}

The amount of vibration is calculated here.
The model pendulum is a 30 cm $\times$ 4 cm $\times$ 3cm bar-shaped mass, whose parameters are listed in Table \ref{table:parameters}.
Two seismic vibration spectrum models are assumed here, which are shown with the blue solid line and dashed line in Fig. \ref{fig:seismic} (a). 
Their respective Trans, Long, Roll and Pitch vibration spectrums along with their resonant modes are shown in Fig. \ref{fig:seismic} (b).
The noisier environment model is close to the measured spectrum in Tokyo, while other less populates places have a similar level as the quiet model. 
Seismic vibration spectrums at several sites are summarized in \cite{RevModPhys.86.121}.
The phase of the seismic vibration at each frequency is assumed to be random, ${\it i.e.}$ the vibration at different frequencies are uncorrelated to each other.

\begin{table}
	\caption{The parameters used for the calculation.}
	\begin{center}
	\begin{tabular}{lllccl}\hline\hline
		parameter		&&			symbol		&	value				&	unit	\\ \hline
		mass			&&			$m$			&	1.0					& 	kg	 	\\ 
		moment of inertia	&(Pitch)&	$I_{\rm P}$	&	$0.23\times10^{-3}$	&	kg$\cdot$m$^2$	\\
						&(Roll)&	 	$I_{\rm R}$	&	$7.64\times10^{-3}$	&	kg$\cdot$m$^2$	\\
						&(Yaw)&		$I_{\rm Y}$	&	$7.58\times10^{-3}$	&	kg$\cdot$m$^2$	\\
		length of wire		&&			$l$			&	0.3					&	m		\\
		suspension point height	&&	$h$			&	0.005				&	0.15			&	m		\\
		Q factors		&&	$Q_x,Q_y,Q_{\rm P},Q_{\rm R}$		& 	$10^3$			&	-		\\
		resonant frequency& translation (x) & $f_x$	&	0.912 				&	Hz		\\
						& translation (y) & $f_y$	&	0.901				&	Hz		\\
						& Pitch	 & 	$f_{\rm P}$	&	2.469				&	Hz		\\
						& Roll	 & 	$f_{\rm R}$	&	0.405				&	Hz		\\	
		\hline\hline
	\end{tabular}
	\label{table:parameters}
	\end{center}
\end{table}

\begin{figure}
\begin{center}
\includegraphics[width=10cm]{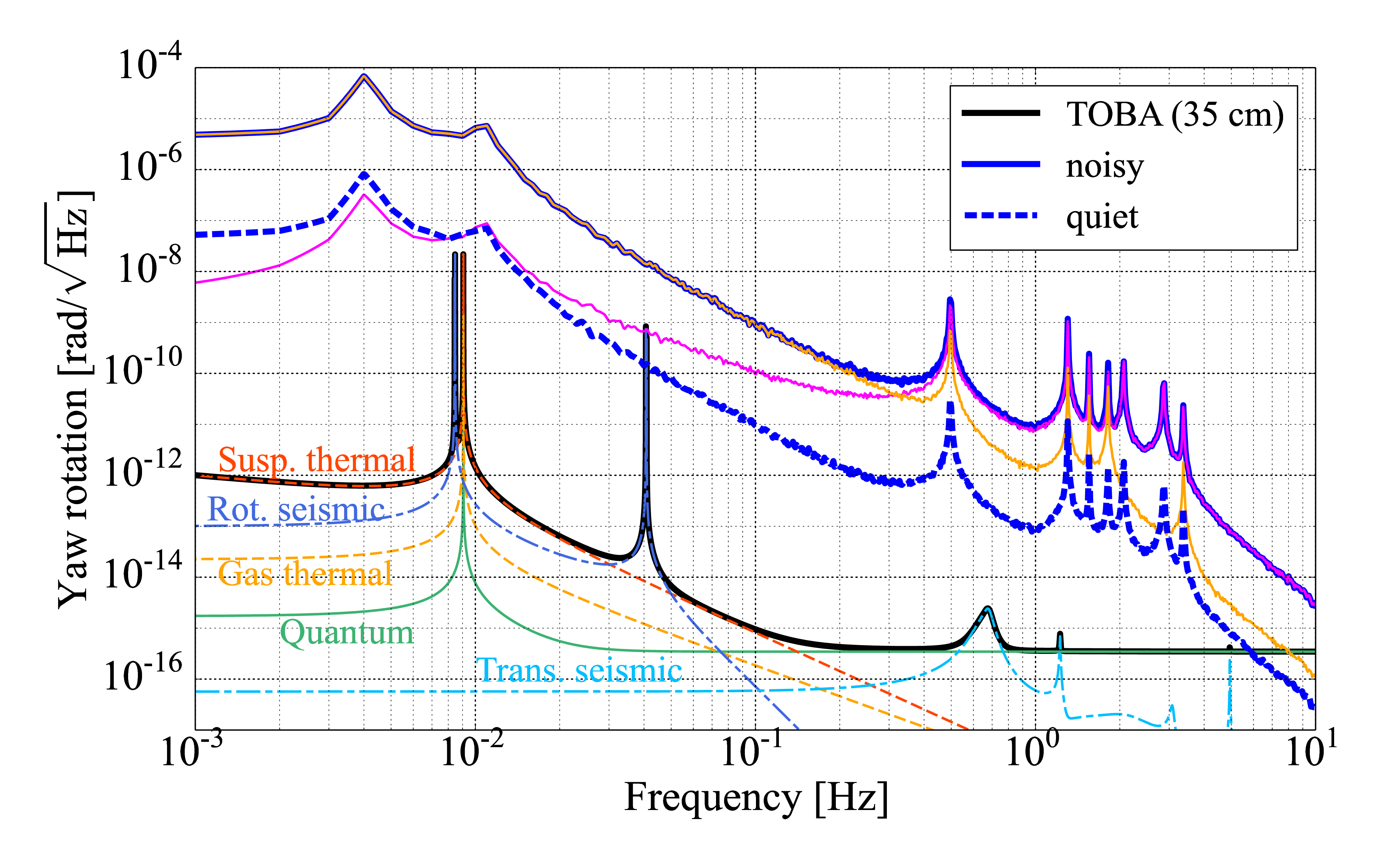}
\caption{\label{fig:calculation} The calculated amplitude spectral density of nonlinearly induced Yaw rotation for large vibration case (solid blue line) and quiet case (dashed blue line). For the case of noisy ground vibration, the solid pink line is the sum of the first three terms of Eq. (\ref{eq:theta_Y}), and the solid orange line shows the last term of Eq. (\ref{eq:theta_Y}). The target sensitivity of TOBA with 35 cm bars is shown with the solid black line, with the noise sources shown in Fig. \ref{fig:target} (b).}
\end{center}
\end{figure}

The calculation results from Eq. (\ref{eq:theta_Y}) for the two seismic vibration models are shown in Fig. \ref{fig:calculation} with blue lines.
The contribution from the first three terms of Eq. (\ref{eq:theta_Y}), which are related to the moments of inertia, is drawn with the pink line.
The last term of Eq. (\ref{eq:theta_Y}), which is proportional to the mass, is shown with the orange line.
They are dominant at higher and lower frequencies, respectively, and their cross-over frequency is 0.3 Hz in this case.

Total torsion angle noise is about $10^{-9}$ rad/$\sqrt{\rm Hz}$ for the seismically noisy case and $10^{-11}$ rad/$\sqrt{\rm Hz}$ for the quiet case at 0.1 Hz.
In terms of torque, they correspond to $3\times10^{-11} \, {\rm N}\cdot{\rm m}/\sqrt{\rm Hz}$ and $3\times10^{-13} \, {\rm N}\cdot{\rm m}/\sqrt{\rm Hz}$, respectively.
Since the nonlinear Yaw rotation originates from the convolution of two degrees of freedom, reduction of seismic vibration by one order of magnitude results in a two orders of magnitude suppression of nonlinear noise. 
In any case, nonlinear noise is much higher than the other noise sources and cannot be ignored at the target sensitivity as a gravity gradiometer, which is at least $10^{-15}$ rad/$\sqrt{\rm Hz}$ at 0.1 Hz.
Therefore we need a strategy to suppress the nonlinear noise, which will be discussed in the next section.

The amplitude spectral density (ASD) in Fig. \ref{fig:calculation} has many peaks.
The frequencies of these peaks correspond to the sum or the difference of the resonant frequencies of the translational modes in $x$, $y$, Roll and Pitch modes, since the Yaw rotation is induced by the frequency convolutions of these degrees of freedom.
For example, the peak at 0.5 Hz originates from the 0.9 Hz resonance of the translational mode in $y$ and the 0.4 Hz resonance of the Roll rotational mode, which are convolved via the last term of Eq. (\ref{eq:theta_Y}).

\section{Reduction of nonlinear vibration noise}\label{sec:reduction}
In this section we discuss how to reduce the nonlinear vibration noise.
The discussion is based on Eq. (\ref{eq:theta_Y}).
To do so, we first give an analytic formula of the frequency convolution with some approximations.

\begin{figure}
\begin{center}
\includegraphics[width=10cm]{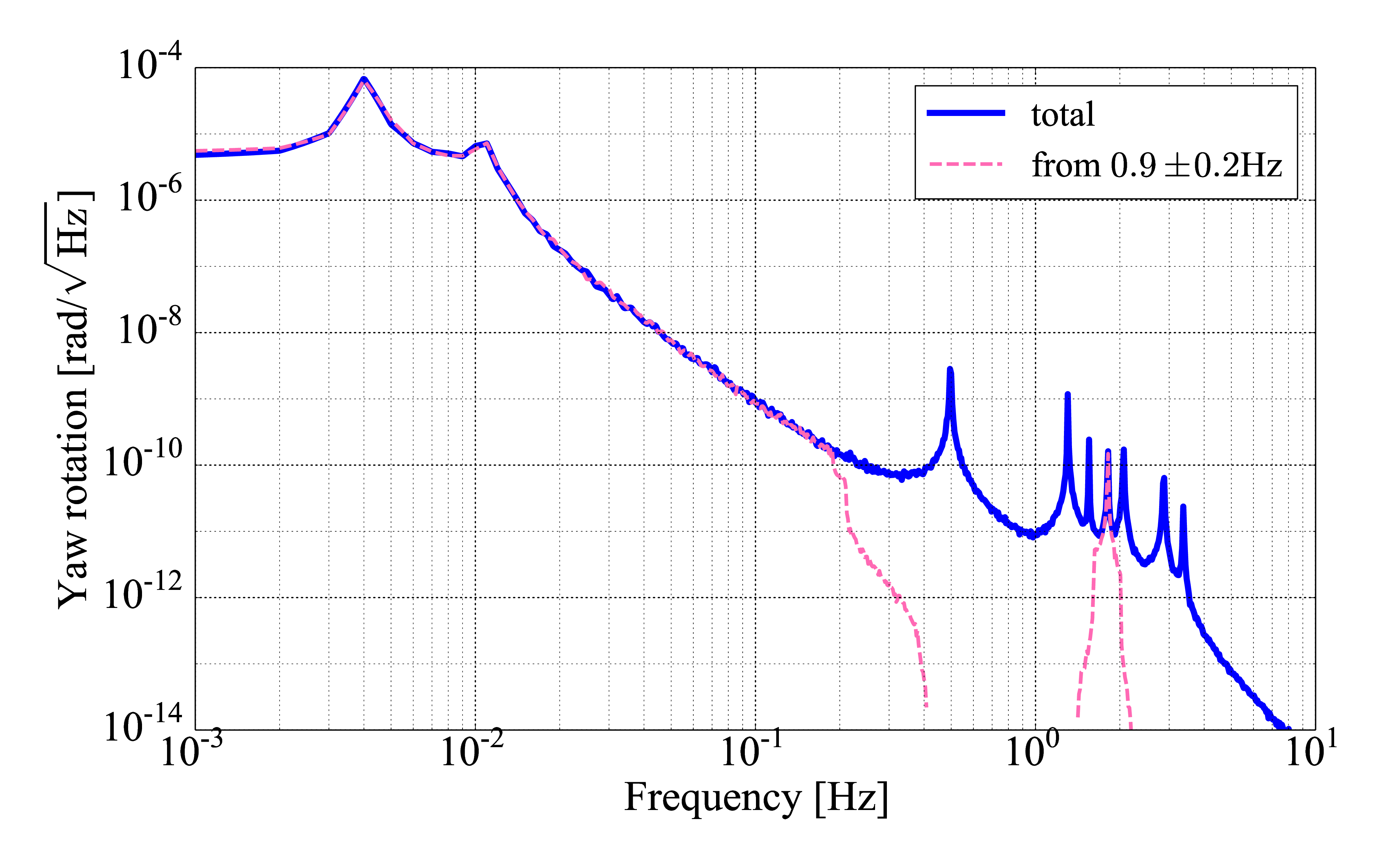}
\caption{\label{fig:aroundf0} The calculated nonlinear noise (solid blue, same as Fig. \ref{fig:calculation}) and the contribution from around the translational resonance peaks, with $\tilde{x}$, $\tilde{y}$, $\tilde{\theta}_{\rm P}$ and $\tilde{\theta}_{\rm R}$ limited to $0.9\pm0.2$ Hz (dashed pink).}
\end{center}
\end{figure}

The dashed pink line in Fig. \ref{fig:aroundf0} is the contribution from the frequency range around the translational resonance (0.9 Hz), which is calculated from Eq. (\ref{eq:theta_Y}) by limiting the frequency range of $\tilde{x}$, $\tilde{y}$, $\tilde{\theta}_{\rm P}$ and $\tilde{\theta}_{\rm R}$ to $0.9\pm0.2$ Hz.
It has a dominant contribution to the total ASD below 0.2 Hz. 
This is because the resonant frequencies of $x$ or $\theta_{\rm R}$ and of $y$ or $\theta_{\rm P}$ are very close (0.912 Hz and 0.901 Hz) so that the convolution between their peaks becomes large.
Therefore, total nonlinear noise can be approximated by the convolution between these resonant peaks.
Additionally, as shown in Fig. \ref{fig:calculation}, the terms of $\left( \omega^2 \tilde{x} \right) \ast \tilde{\theta}_{\rm P}$ and $\left( \omega^2 \tilde{y} \right) \ast \tilde{\theta}_{\rm R}$ of Eq. (\ref{eq:theta_Y}) are dominant below 0.3 Hz.
Since Pitch rotation is larger than Roll rotation around 0.9 Hz (Fig. \ref{fig:seismic} (b)), $\left( \omega^2 \tilde{x} \right) \ast \tilde{\theta}_{\rm P}$ is the most important term for the approximation.

\begin{figure}
\begin{center}
\includegraphics[width=10cm]{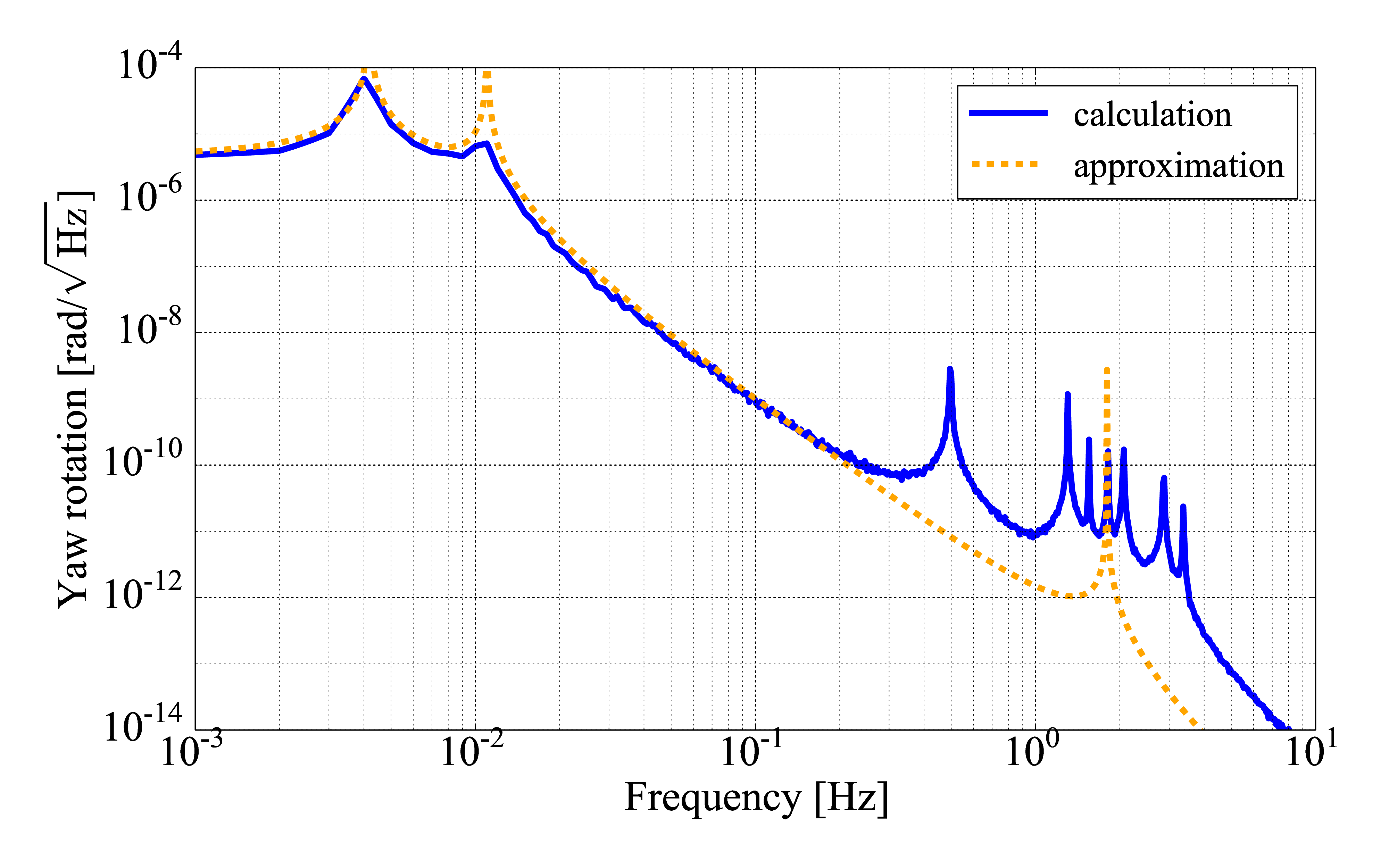}
\caption{\label{fig:approx} The approximated formula Eq. (\ref{eq:Yaw_approx}) (dotted orange) and the full calculation result (solid blue). The approximation is close to the full calculation below 0.3 Hz.}
\end{center}
\end{figure}

The amplitude spectral density of the convolution term $\left( \omega^2 \tilde{x} \right) \ast \tilde{\theta}_{\rm P}$ is calculated in  \ref{sec:appendix}.
From Eq. (\ref{eq:G_xP}) and (\ref{eq:theta_Y}), the approximated formula of the ASD of $\theta_{\rm Y}$ is
\begin{eqnarray}
\fl \hspace{10pt}
\sqrt{G_{\theta_{\rm Y}}(f)} &\approx \left| \frac{mh}{\kappa_{\rm Y} - I_{\rm Y} (2\pi f)^2} \right| \sqrt{G_{\left( \omega^2 \tilde{x} \right) \ast \tilde{\theta}_{\rm P}}(f)} \nonumber\\
\fl
&= \left| \frac{mh}{\kappa_{\rm Y} - I_{\rm Y} (2\pi f)^2} \right| \frac{(2\pi \times 1\,{\rm Hz})^4}{g} G_{\rm seis}({1\,\rm Hz}) \nonumber\\\
\fl
&\hspace{100pt}
\times
\sqrt{ \frac{\pi Q f_0{}^5 }{(f-|f_x-f_y|)^2} \left( \frac{1}{(f-2f_0)^2} + \frac{1}{(f+2f_0)^2} \right) }. \label{eq:Yaw_approx}
\end{eqnarray}
This formula is valid at low frequencies (below $\sim0.3$ Hz).
Fig. \ref{fig:approx} compares this approximation and the full calculation (Fig. \ref{fig:calculation}).
It shows that Eq. (\ref{eq:Yaw_approx}) gives a good approximation below 0.3 Hz.
Several dependences on the parameters can be extracted from this formula, which are discussed below.

\subsection{Vibration isolation around resonant frequencies}
We have already shown that reduction of seismic vibration is effective for suppression of nonlinear noise.
In particular, vibration isolation around the pendulum's resonant frequencies, which typically lie around 1 Hz for a cm -- m scale pendulum, is important for broadband noise suppression.
This is because the large fraction of the frequency convolution originate from the vibrations at resonant frequencies convolved with the nearby frequencies.

Several vibration isolation systems around 1 Hz or sub-Hz have been proposed.
Note that when using them, we have to be careful about additional resonances, which can induce additional nonlinear noise.
An active feedback system with a hexapod stage will be a good choice for this reason, because it uses a rigid system so that the additional resonant frequencies are high enough. 
Vibration isolation of about one order of magnitude with the hexapod system at 0.5 -- 5 Hz is reported in \cite{UshibaThesis}.

\begin{figure}
\begin{center}
	\begin{tabular}{ll}
	\begin{minipage}{0.45\hsize}
	\begin{center}
	\includegraphics[width=7cm]{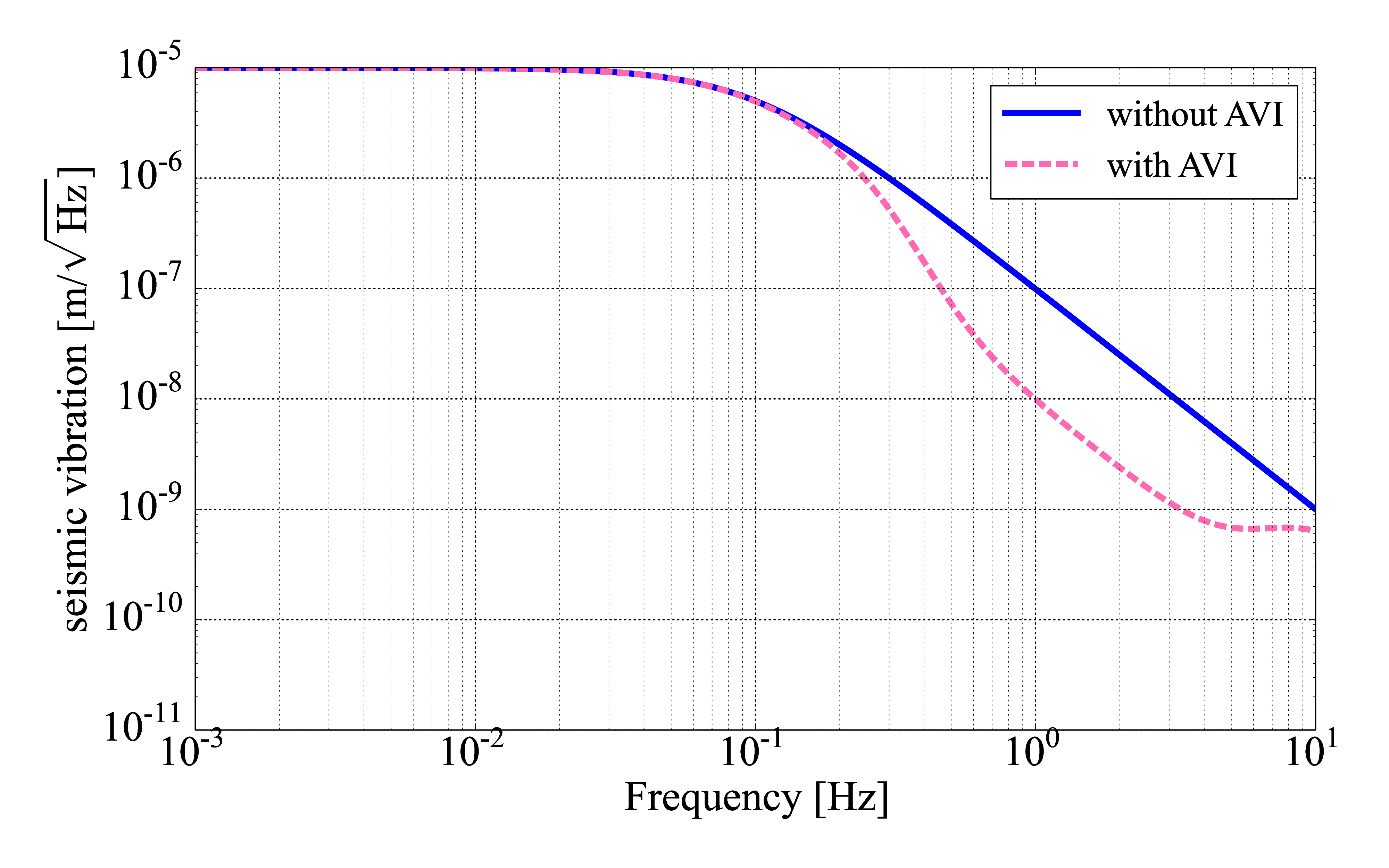}
	(a)
	\end{center}
	\end{minipage}
	\begin{minipage}{0.45\hsize}
	\begin{center}
	\includegraphics[width=7cm]{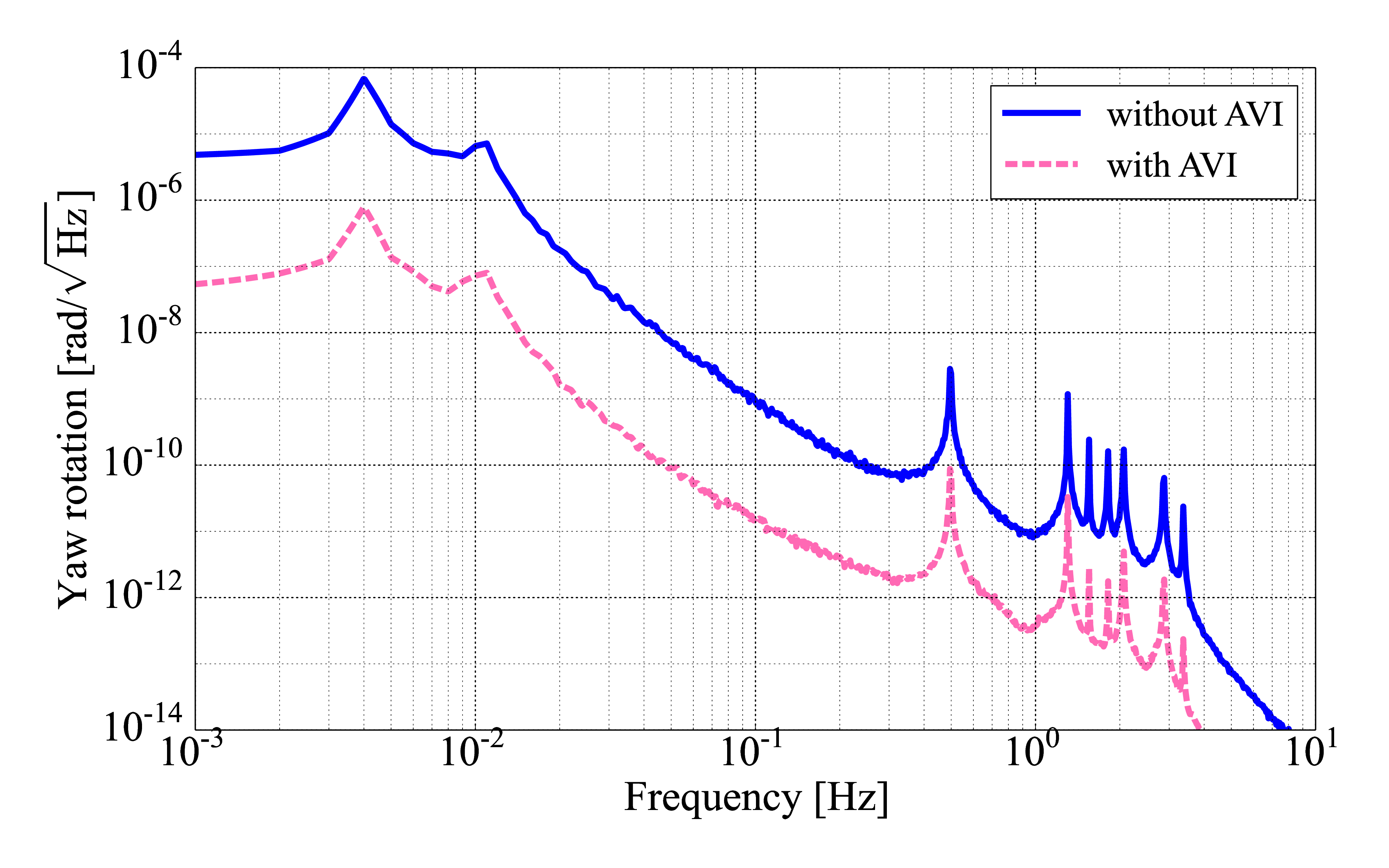}
	(b)
	\end{center}
	\end{minipage}
	\end{tabular}
\caption{\label{fig:isolation}The amplitude spectral densities of (a) the seismic vibration and (b) nonlinear vibration noise with (dashed pink) and without (solid blue) active vibration isolation (AVI).}
\end{center}
\end{figure}

The calculated nonlinear noise with active vibration isolation (AVI) is shown in Fig. \ref{fig:isolation}.
The assumed vibration is shown in Fig. \ref{fig:isolation} (a), which is the original vibration suppressed by one order of magnitude at 0.5 -- 5 Hz.
AVI successfully reduces nonlinear noise by two orders of magnitudes.
At every frequency, it reaches almost the same noise level as the quiet case in Fig. \ref{fig:calculation}, which confirms the importance of isolation around the pendulum's resonant frequencies.

\subsection{Suppressing resonances}
For similar reasons as the previous discussion, suppressing the resonant peaks by damping is also effective in reducing nonlinear noise.
Although the Q factors of various non-torsional resonant modes do not appear explicitly in Eq. (\ref{eq:theta_Y}), the convolution terms of Eq. (\ref{eq:theta_Y}) are dependent on them.
As shown in Eq. (\ref{eq:Yaw_approx}), nonlinear noise is proportional to $\sqrt{Q}$.
This is because the RMS amplitude of a resonant mode is proportional to $\sqrt{Q}$.

\begin{figure}
\begin{center}
\includegraphics[width=10cm]{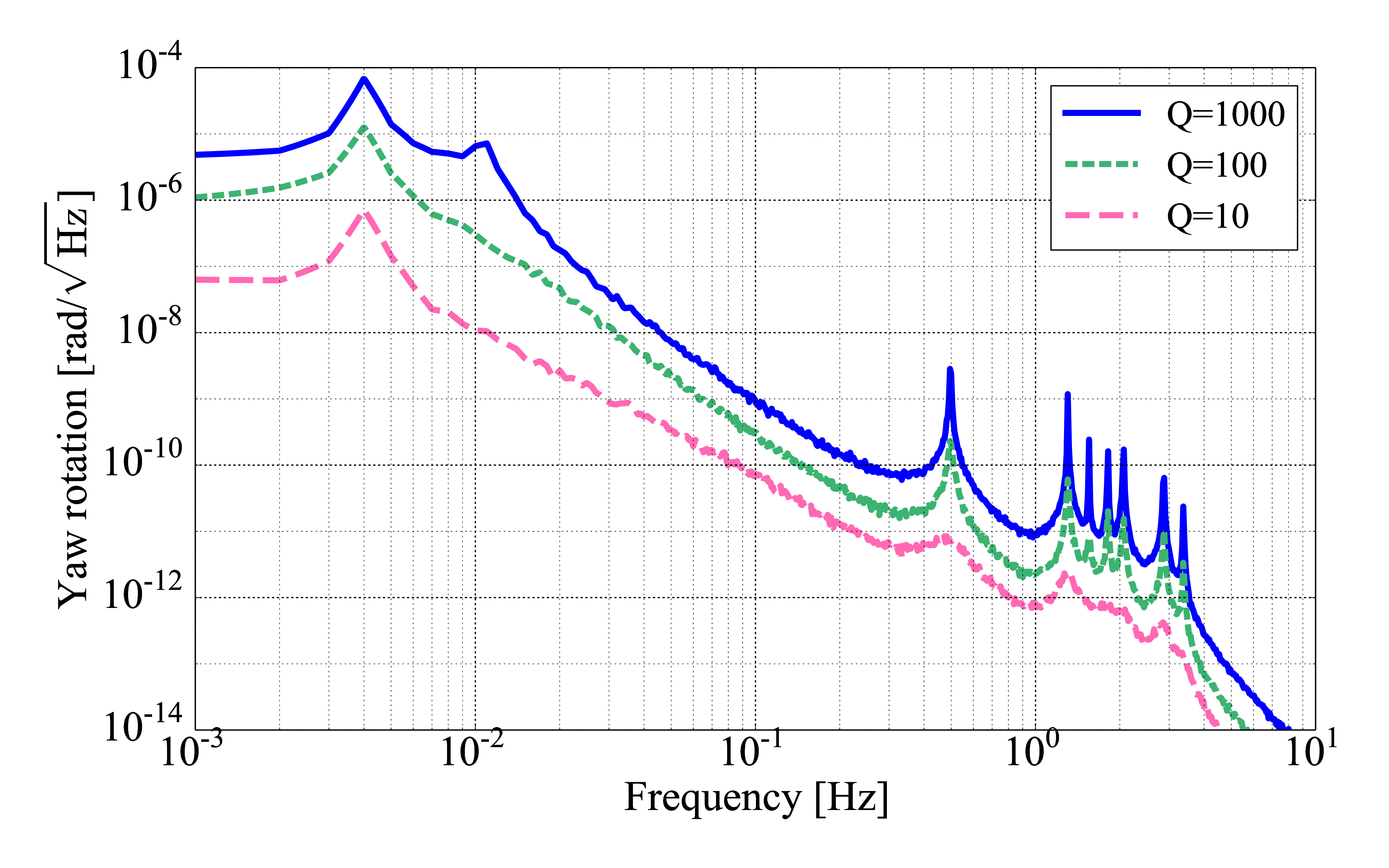}
\caption{\label{fig:Q} Q factor dependence of nonlinear vibration noise. $Q=10^3$ (solid blue), $Q=10^2$ (dashed green) and $Q=10$ (long-dashed pink).}
\end{center}
\end{figure}

Fig. \ref{fig:Q} shows the calculated nonlinear noise with $Q=10^3$, $10^2$ and $10$.
All resonant peaks are assumed to have the same Q factor.
As expected, lower Q gives smaller noise, and the dependence is almost proportional to $\sqrt{Q}$.
This shows the importance of damping in terms of noise.

\subsection{Proper choice of pendulum's parameters}
Parameters of the pendulum are also important factors of nonlinear noise.
Some of them appear explicitly in Eq. (\ref{eq:theta_Y}), and some are also implicitly related to $\tilde{x}$, $\tilde{y}$, $\tilde{\theta}_{\rm R}$ and $\tilde{\theta}_{\rm P}$ via their resonant frequencies.
Additionally, they are technically related to the achievable Q factors with some damping systems.
The Q factors are not in Eq. (\ref{eq:theta_Y}) explicitly, but they are related to $\tilde{x}$, $\tilde{y}$, $\tilde{\theta}_{\rm R}$ and $\tilde{\theta}_{\rm P}$, and approximately appears as Eq. (\ref{eq:Yaw_approx}).
Since these issues are correlated, dependence on a parameter is a complicated question.
Here we show some examples of dependence, though a complete discussion will require more consideration with a detailed set of conditions.

First we investigate the resonant frequencies.
Resonant frequencies are the key factors in determining the vibration transfer function of the pendulum.
As Eq. (\ref{eq:Yaw_approx}) shows, nonlinear noise will be proportional to $f_0{}^{1.5}$ at low frequencies.

\begin{figure}
\begin{center}
\includegraphics[width=10cm]{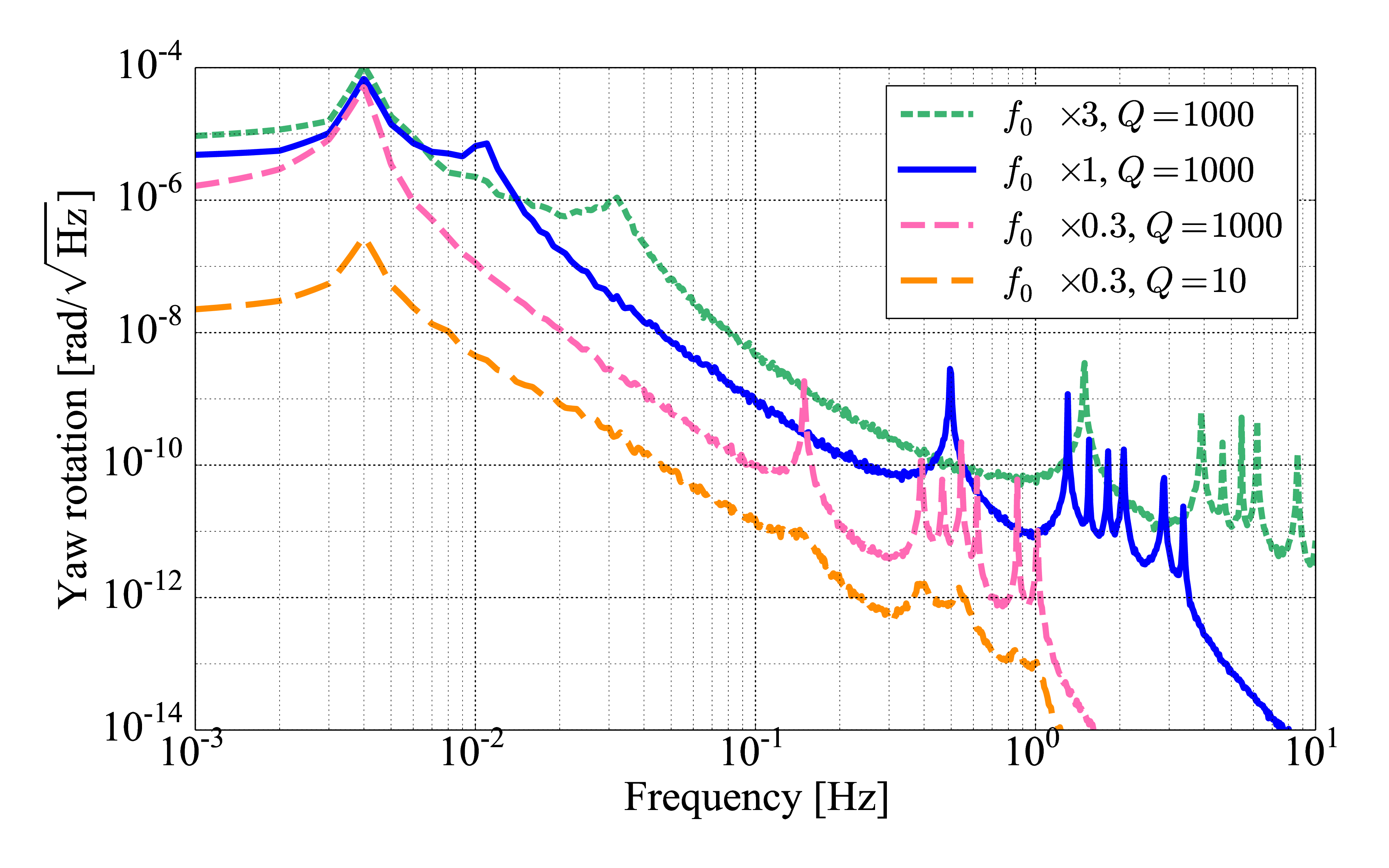}
\caption{\label{fig:f0} Dependence on resonant frequencies. All the resonant frequencies are scaled by $\times3$ (dashed green), $\times1$ (solid blue) and $\times1/3$ (long-dahsed pink) in case of $Q=1000$. Additionally, the longer-dashed orange line shows the nonlinear noise when the resonant frequencies are scaled by $\times1/3$ and the Q factor is suppressed to 10.}
\end{center}
\end{figure}

Consider the case where all the resonant frequencies are scaled by the same factor while the other parameters, such as $h$ or the size of the suspended bar, are fixed.
Though this assumption is a little unrealistic because resonant frequencies are determined by the parameters, we want to extract only the dependence on the resonant frequencies here.
In this case, nonlinear noise will change as in Fig. \ref{fig:f0}.
The dependence on the resonant frequencies is roughly as expected at low frequencies.
Therefore we should set resonant frequencies lower.

Note that this simple dependence is true only when the seismic vibration spectrum is a simple function of frequency.
The actual spectrum has some structure as shown in Fig. \ref{fig:seismic}, so there may be an optimal choice of resonant frequencies.

Suppression of the Q factor gives an additional reduction of nonlinear noise in proportion to $\sqrt{Q}$ (long-dashed orange line in Fig. \ref{fig:f0}).
The same dependence on the Q factor as shown in Fig. \ref{fig:Q} is true even if the resonant frequencies are different.

\begin{figure}
\begin{center}
\includegraphics[width=10cm]{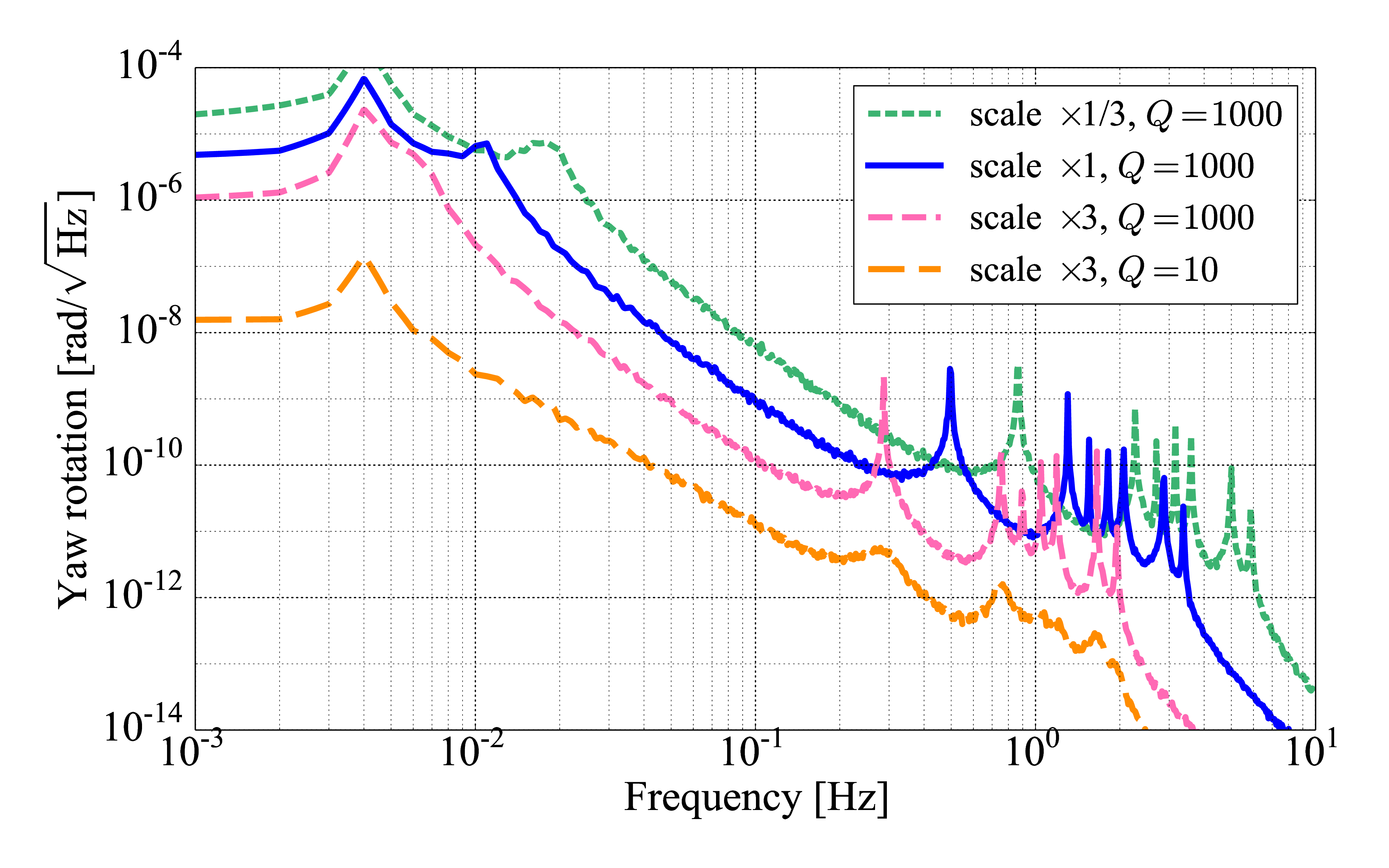}
\caption{\label{fig:scale} Nonlinear noise when the pendulum scales by 1/3 (dashed green), 1 (solid blue) and 3 (long-dashed pink). The longer-dashed orange line shows the nonlinear noise when the pendulum scales by $\times3$ and the Q factor is suppressed to 10.}
\end{center}
\end{figure}

Next we discuss the size of the pendulum.
When every length ($l, \, h$, and size) scales by the same factor, $1/3$, $1$ or $3$, nonlinear noise will change as seen in Fig. \ref{fig:scale}.
It shows a rough proportionality to the inverse-square of the scaling factor.
This is partly because the dominant term of Eq. (\ref{eq:theta_Y}) is proportional to $mh/I_{\rm Y}$, which in turn is proportional to the inverse of the scaling factor.
The change of resonant frequencies, which was discussed above, also contributes.
The resonant frequencies are given by Eq. (\ref{eq:resonances}), so all of them are proportional to the inverse square-root of the scaling factor.
The total dependence of noise can be explained by these two factors.

Additional reduction can be achieved by reducing the Q factor in conjunction as shown in Fig. \ref{fig:scale} (long-dashed orange line).


\section{Discussion}\label{sec:discussion}
Taking into account nonlinear vibration transfer in the mechanical system, sensitivity of a 30 cm scale pendulum is limited to $10^{-9}$ rad/$\sqrt{\rm Hz}$ at 0.1 Hz (Fig. \ref{fig:calculation}) in a noisy environment like Tokyo without any active vibration isolation ($\sim 10^{-7}$ m/$\sqrt{\rm Hz}$ at 1 Hz (Fig. \ref{fig:seismic})) or passive damping of the other pendulum modes ($Q=1000$). 
In terms of torque, this corresponds to $3\times10^{-11}$ N$\cdot$m/$\sqrt{\rm Hz}$.
These values are much larger than the target sensitivity for the usage of gravity gradient measurement, which is $10^{-15}$ rad/$\sqrt{\rm Hz}$ at 0.1 Hz for EEW, and $10^{-19}$ rad/$\sqrt{\rm Hz}$ at 0.1 Hz for GW observation.
Hence we need a strategy to suppress nonlinear noise by at least six orders of magnitude at 0.1 Hz.

\begin{figure}
\begin{center}
\includegraphics[width=10cm]{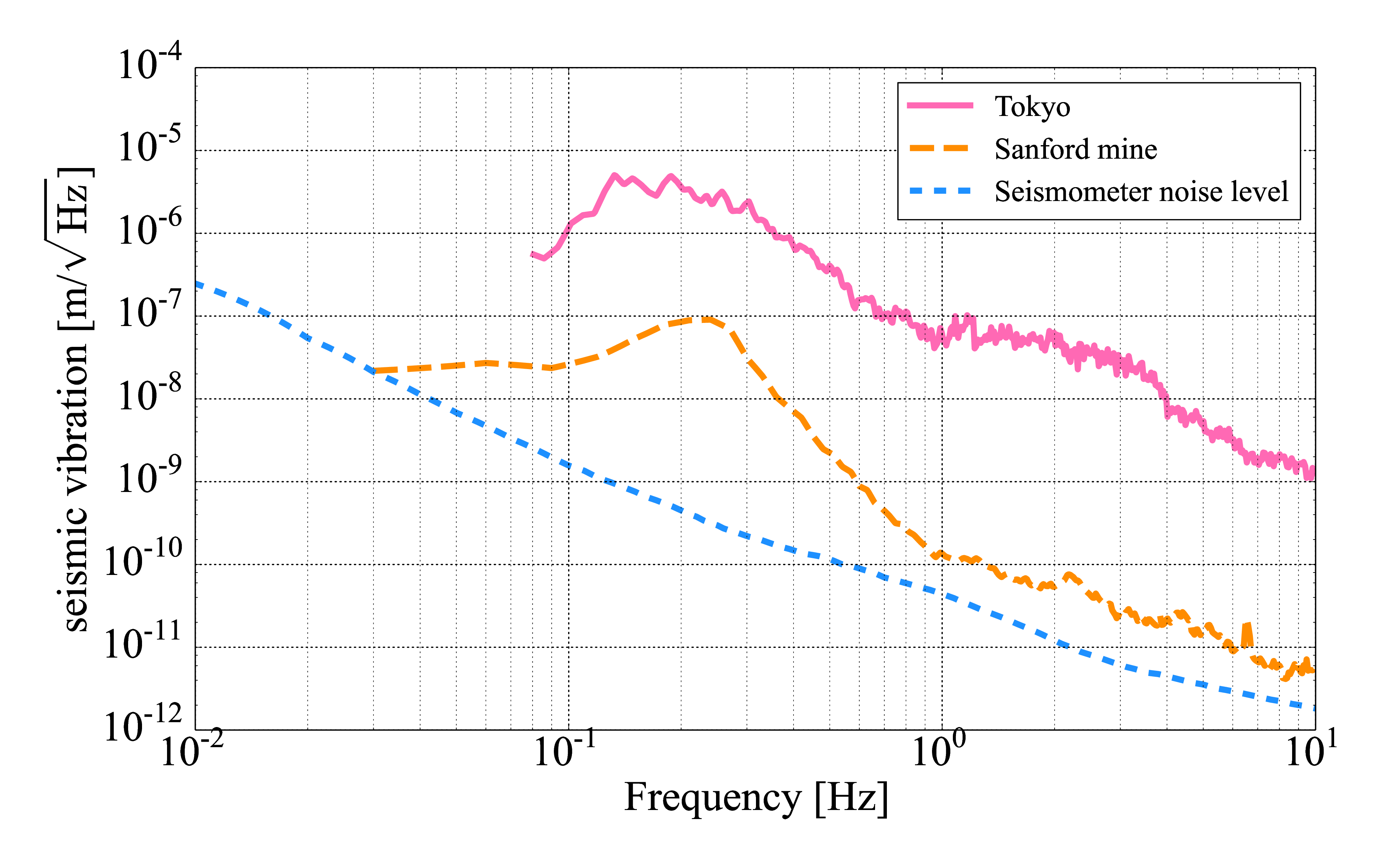}
\caption{\label{fig:seismic2} The ASD of seismic vibration at Sanford mine in the U.S. (dashed orange) \cite{RevModPhys.86.121} and the instrumental self-noise level of a low-noise seismometer (dotted blue) \cite{AVI}, which gives the achievable vibration level with an active vibration isolation.}
\end{center}
\end{figure}

\begin{figure}
\begin{center}
\includegraphics[width=10cm]{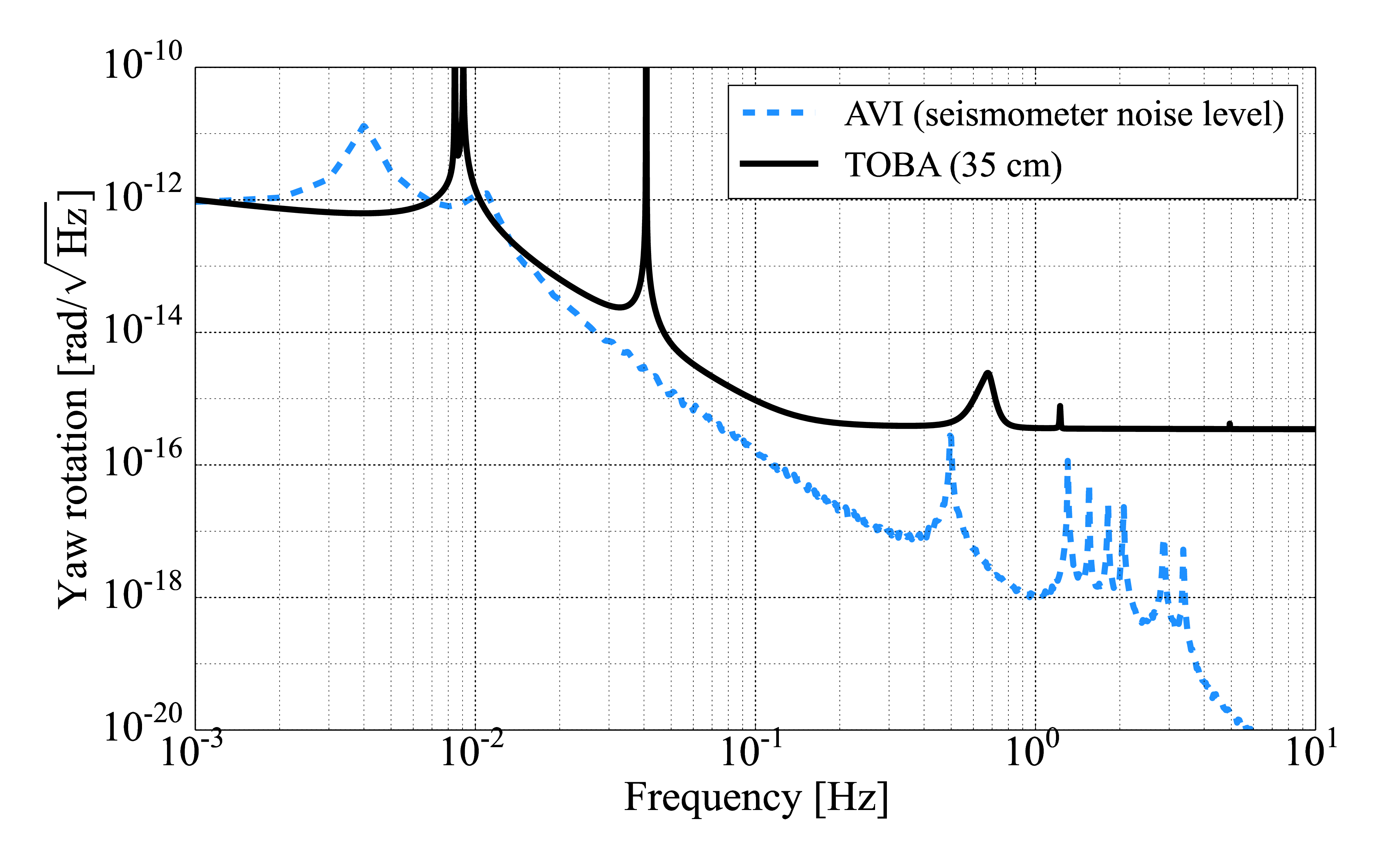}
\caption{\label{fig:reduction_phase3} The achievable nonlinear noise level for the 30 cm scale pendulum by actively suppressing the vibration to the noise level of the low-noise seismometer (dotted blue). The Q factors are set to 1000.}
\end{center}
\end{figure}

As shown in Fig. \ref{fig:isolation} (b), vibration isolation around the resonant frequency by one order of magnitude reduces nonlinear noise by two orders of magnitude.
Hence if the vibration of the top suspension point is suppressed by two more orders of magnitudes, to $10^{-10}$ m/$\sqrt{\rm Hz}$ around 1 Hz, nonlinear noise of a 30 cm scale pendulum will be reduced by four more orders of magnitude and reach $10^{-15}$ rad/$\sqrt{\rm Hz}$ at 0.1 Hz.
Fig. \ref{fig:seismic2} shows the ASD of the noise level of low-noise seismometers \cite{AVI}, which is below $10^{-10}$ m/$\sqrt{\rm Hz}$ around 1 Hz.
Hence an active vibration isolation system with seismometers can suppress the vibration down to this noise level in principle.
By using such a system, nonlinear noise for the 30 cm scale pendulum is reduced to below $10^{-15}$ rad/$\sqrt{\rm Hz}$ at 0.1 Hz as shown in Fig. \ref{fig:reduction_phase3}, which satisfies the requirement for EEW.
Note that careful design is required to realize the active vibration isolation system.
In particular, tilt-horizontal coupling can be a problem \cite{THcoupling}, so some form of tiltmeter for the decoupling may be necessary.
Even if it is technically difficult, additional noise reduction by damping or tuning of the resonant frequency can ease the requirement.
For example, when the resonances are damped to $Q=10$, the vibration of the top suspension point can be $\sqrt{10}\sim 3$ times larger than the value above (Fig. \ref{fig:Q}).

If we choose a seismically quiet site to build the detector, nonlinear noise can be reduced more easily.
It is reported that the vibration at Sanford mine in the U.S. is $10^{-10}$ m/$\sqrt{\rm Hz}$ at 1 Hz \cite{RevModPhys.86.121}.
The ASD is shown in Fig. \ref{fig:seismic2} with the seismic vibration in Tokyo and the noise level of low-noise seismometers.
However, note that we may not always have an arbitrary choice of site for the purpose of earthquake early warning.
Though optimal arrangement of sensors is under investigation, some of the detectors will be placed near the city the system is meant to protect, where vibration tends to be large.
Therefore, designing an isolation system is necessary.

\begin{figure}
\begin{center}
\includegraphics[width=10cm]{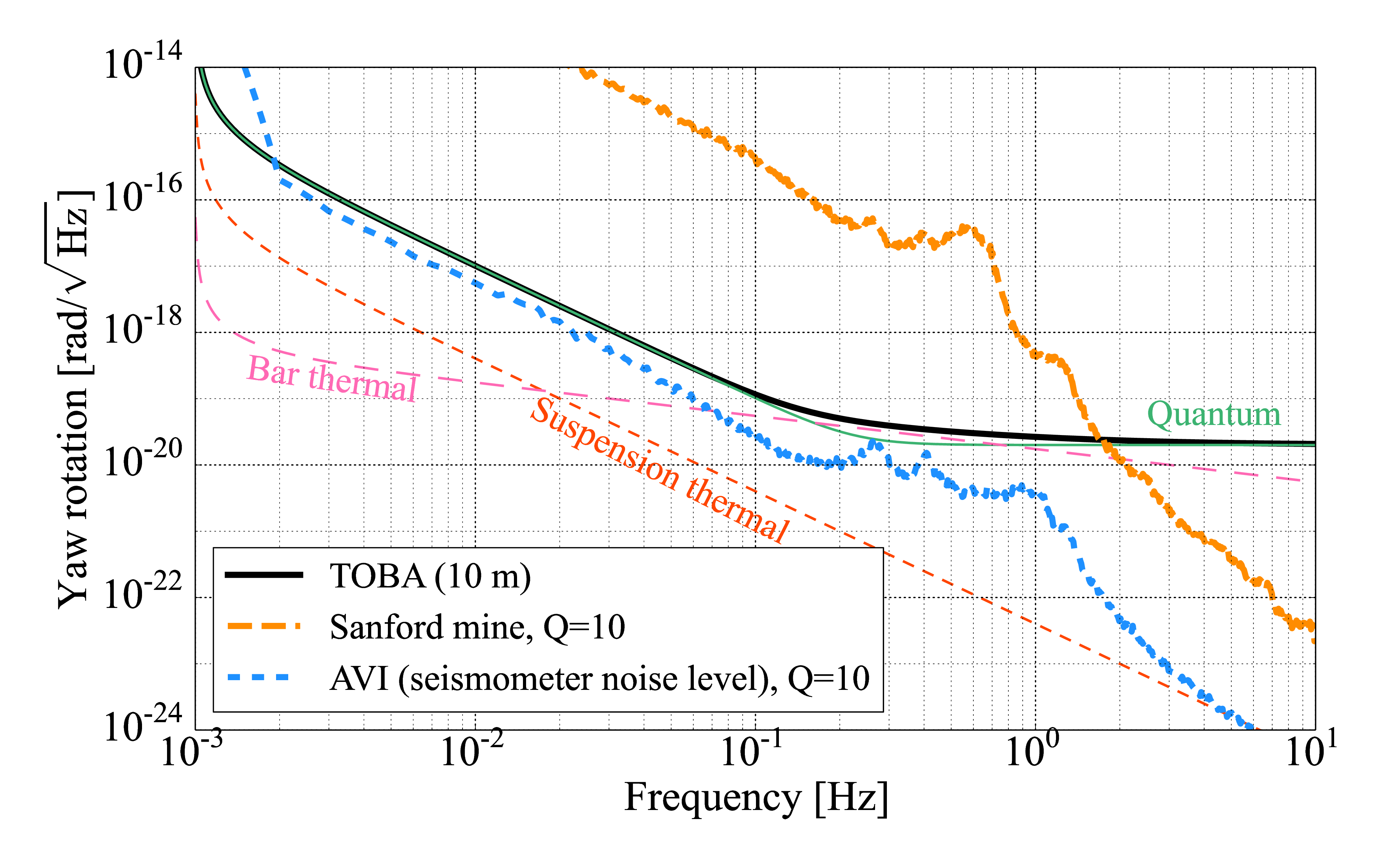}
\caption{\label{fig:final} The achievable nonlinear noise level for the 10 m scale pendulum by placing the detector at Sanford mine (dashed orange) or actively suppressing the vibration to the noise level of the low-noise seismometer (dotted blue), plotted with other noise sources (solid green : quantum noise, dashed red : suspension thermal noise, long-dashed pink : bar thermal noise \cite{TOBA}). The Q factors are set to 10.}
\end{center}
\end{figure}

For gravitational wave observations, 10 m scale pendulums are planned \cite{TOBA}.
The parameters are listed in Table \ref{table:final}. 
Since the design of the wire length and the suspension point height is not fixed yet, arbitral values of $l=2$ m and $h=0.15$ m are set for them here.
By applying the scale dependence which was discussed above in Fig. \ref{fig:scale}, we can expect three orders of magnitude noise suppression compared with 30 cm.
Additionally, the active vibration isolation using the low-noise seismometers can suppress the vibration by 3.5 orders of magnitude in Tokyo (Fig. \ref{fig:seismic2}), which will reduce nonlinear noise by seven orders of magnitude more.
Damping to $Q=10$ gives an additional reduction of one order of magnitude.
In total, eleven orders of magnitude reduction can be expected, which results in nonlinear noise of $10^{-20}$ rad/$\sqrt{\rm Hz}$ at 0.1 Hz.
Though we can choose seismically quiet sites such as Sanford mine to build in for the lower level of seismic vibration, the advantage of Sanford mine is not significant since the resonant frequencies are lower than the 30 cm scale pendulum as listed in Table \ref{table:final}.
The difference of the vibration ASD between Tokyo and Sanford mine is only 1.5 orders of magnitude at the resonant frequencies (sub-Hz) (Fig. \ref{fig:seismic2}), so only three orders of magnitude suppression is expected for nonlinear noise.
Therefore active vibration isolation is essential again for this case.
Fig. \ref{fig:final} shows nonlinear noise of a 10 m scale pendulum if it is placed at Sanford mine or the vibration is actively suppressed to the noise level of a seismometer. 
$10^{-19}$ rad/$\sqrt{\rm Hz}$ at 0.1 Hz, the level of quantum noise and thermal noise of the bar, is shown to be achievable by combining the active vibration isolation system with damping.

\begin{table}
	\caption{The parameters of 10 m scale TOBA.}
	\begin{center}
	\begin{tabular}{lllcl}\hline\hline
		parameter		&&			symbol		&	value				&	unit	\\ \hline
		mass			&&			$m$			&	7600				& 	kg	 	\\ 
		moment of inertia	&(Pitch)&	$I_{\rm P}$	&	$6.4\times10^{4}$	&	kg$\cdot$m$^2$		\\
						&(Roll)&	 	$I_{\rm R}$	&	$3.4\times10^{2}$	&	kg$\cdot$m$^2$		\\
						&(Yaw)&		$I_{\rm Y}$	&	$6.4\times10^{4}$	&	kg$\cdot$m$^2$		\\
		length of wire		&&			$l$			&	2					&	m		\\
		suspension point height	&&	$h$			&	0.15					&	m		\\
		Q factors		&&	$Q_x,Q_y,Q_{\rm P},Q_{\rm R}$	& 	$10$		&	-		\\
		resonant frequency& translation (x) & $f_x$	&	0.353 				&	Hz		\\
						& translation (y) & $f_y$	&	0.337				&	Hz		\\
						& Pitch	 & 	$f_{\rm P}$	&	0.952				&	Hz		\\
						& Roll	 & 	$f_{\rm R}$	&	0.066				&	Hz		\\
		\hline\hline
	\end{tabular}
	\label{table:final}
	\end{center}
\end{table}

These rough plans give us a prospect to reach the target sensitivities as a gravity gradiometer.
For more strict calculations, however, detailed design of the suspension system and the structure of seismic vibration spectrum have to be taken into account.
Though we do not go into the details of this here, it will be necessary when doing case studies.


\section{Conclusion}
We have discussed about how the rotational noise of a torsion pendulum is introduced nonlinearly from seismic vibrations.
The ASD of the noise can reach $10^{-9}$ rad/$\sqrt{\rm Hz}$ at 0.1 Hz for a 30 cm scale pendulum in locations with noisy seismic vibrations such as Tokyo.
Based on the investigation in Sec. \ref{sec:reduction}, we have created a rough strategy to reduce the noise down to the target sensitivity, $10^{-15}$ rad/$\sqrt{\rm Hz}$ at 0.1 Hz for EEW and $10^{-19}$ rad/$\sqrt{\rm Hz}$ for GW at 0.1 Hz.
They are achievable in principle by combining the following; active vibration isolation, damping system, increasing the size of the pendulum or going to seismically quiet site.
Note that a strict estimation of noise will require a case study based on the detailed design of the suspension system and the actual seismic vibration spectrum.

\section*{Acknowledgments}
The work about TOBA is supported by JSPS KAKENHI Grants Number JP16H03972, JP24244031 and JP18684005.
We thank Ooi Ching Pin for editing this paper.

\appendix
\section{Convolution of the translation $\tilde{x}$ and the rotation $\tilde{\theta}_{\rm P}$} \label{sec:appendix}
The frequency convolution $\left( \omega^2 \tilde{x}\right)\ast\tilde{\theta}_{\rm P}$ is calculated in this appendix.

First, based on Eq. (\ref{eq:x_fourier}) and (\ref{eq:P_fourier}), the Fourier spectrum of $\tilde{x}$ and $\tilde{\theta}_{\rm P}$ can be approximated as follows around the translational resonant frequency $f_x$ or $f_y$, with
\begin{equation}
\tilde{x}(f) \simeq \frac{ f_x{}^2 }{ f_x{}^2 + i\frac{ f_x }{Q_x} f - f^2 } \left( \frac{\rm 1 Hz}{f} \right)^2 \tilde{x}_{\rm g}({\rm 1 Hz}) \label{eq:x_fourier_approx}
\end{equation}
and
\begin{equation}
\tilde{\theta}_{\rm P}(f) \simeq \frac{4\pi^2f_y{}^2 f^2 }{g \left(f_y{}^2 + i \frac{f_y}{Q_y}f - f^2 \right)} \left( \frac{\rm 1 Hz}{f} \right)^2 \tilde{y}_{\rm g}({\rm 1 Hz}).   \label{eq:P_fourier_approx}
\end{equation}
Here the seismic vibration is assumed to be proportional to inverse-square of frequency. 
$\tilde{x}_{\rm g}({\rm 1 Hz})$ and $\tilde{y}_{\rm g}({\rm 1 Hz})$ are the Fourier spectrum of the seismic vibration at 1 Hz.
These are converted from the amplitude spectral density of the seismic vibration $\sqrt{G_{\rm seis}(f)}$ as
\begin{equation}
\tilde{x}_{\rm g}(f) = \sqrt{\frac{T}{8\pi^2}} \sqrt{G_{\rm seis}(f)} e^{i\theta_x(f)}
\end{equation}
and
\begin{equation}
\tilde{y}_{\rm g}(f) = \sqrt{\frac{T}{8\pi^2}} \sqrt{G_{\rm seis}(f)} e^{i\theta_y(f)}.
\end{equation}
$T$ is the time length used for the calculation. 
The dependence on $T$ disappears later when the results are re-converted to ASD.
Here the phase of the Fourier spectrums, $\theta_x(f)$ and $\theta_y(f)$, are assumed to be random at each frequency and independent between $x$ and $y$.
Then the convolution around the translational resonant frequency $f_x\simeq f_y$ is 
\begin{eqnarray}
\fl
\left( \omega^2 \tilde{x} \right) \ast \tilde{\theta}_{\rm P} (f) \nonumber \\
\fl\hspace{20pt}
= \frac{(2\pi)^4 (1\,{\rm Hz})^4}{g} \frac{T}{8\pi^2} G_{\rm seis}({1\,\rm Hz}) \nonumber \\
\fl\hspace{40pt}
\times \int_{f_0-\Delta f}^{f_0+\Delta f} \frac{ f_x{}^2 }{ f_x{}^2 + i\frac{ f_x }{Q_x} (f-\alpha) - (f-\alpha)^2 } \frac{ f_y{}^2 }{ f_y{}^2 + i \frac{f_y}{Q_y}\alpha - \alpha^2} e^{i (\theta_x(f-\alpha)+\theta_{y}(\alpha))} d\alpha.
\end{eqnarray}
Here $f_0=(f_x+f_y)/2\simeq f_x \simeq f_y$, and $\Delta f$ is the frequency range of convolution, which is chosen to cover the resonant peaks of $x$ and $\theta_{\rm P}$ ($f_0=0.9$ Hz and $\Delta f=0.2$ Hz in Fig. \ref{fig:aroundf0}).
The frequency $f$ is limited to within $|f|<\Delta f$.
Since the integrand has large value only around $f_0$, the integral range can be extended to $[0,\infty]$ in approximation. 
Then the integral part can be modified to 
\begin{equation}
({\rm integral}) \approx \int_0^{\infty} A_f(\alpha) e^{i \theta(\alpha)} e^{i \alpha t} d\alpha \label{eq:integral}
\end{equation}
where
\begin{equation}
A_f(\alpha) \equiv \frac{ f_x{}^2 }{ f_x{}^2 + i\frac{ f_x }{Q_x} (f-\alpha) - (f-\alpha)^2 } \frac{ f_y{}^2 }{ f_y{}^2 + i \frac{f_y}{Q_y}\alpha - \alpha^2} 
\end{equation}
and
\begin{equation}
\theta(\alpha) = \theta_x(f-\alpha)+\theta_{y}(\alpha) - \alpha t.
\end{equation}
Due to the randomness of $\theta_x$ and $\theta_y$, $\theta(\alpha)$ is a random phase independent to $t$, so Eq. (\ref{eq:integral}) is a Fourier transform of $A_f(\alpha) e^{i \theta(\alpha)}$.
Therefore, the mean power of $\left( \omega^2 \tilde{x} \right) \ast \tilde{\theta}_{\rm P} (f)$ is 
\begin{eqnarray}
\fl \hspace{20pt}
\left< \left| \left( \omega^2 \tilde{x} \right) \ast \tilde{\theta}_{\rm P} (f) \right|^2 \right> 
&= 
\left( \frac{(2\pi \times 1\,{\rm Hz})^4}{g} \frac{T}{8\pi^2} G_{\rm seis}({1\,\rm Hz}) \right)^2 \left< \left| \mathcal{F.T.} \left[ A_f(\alpha) e^{i \theta(\alpha)} \right] \right|^2 \right> \nonumber\\
&= 
\left( \frac{(2\pi \times 1\,{\rm Hz})^4}{g} \frac{T}{8\pi^2} G_{\rm seis}({1\,\rm Hz}) \right)^2 \int_0^{\infty} G_A(\alpha) d\alpha. \label{eq:xP_power}
\end{eqnarray}
$G_A$ is the power spectral density of $A_f(\alpha) e^{i \theta(\alpha)}$, so
\begin{eqnarray}
\fl \hspace{20pt}
\int_0^{\infty} G_A(\alpha) d\alpha &= \int_0^{\infty} \frac{8\pi^2}{T} \left| A_f(\alpha) e^{i \theta(\alpha)} \right|^2 \nonumber \\
&= \frac{8\pi^2}{T} \int_0^{\infty} \frac{f_x{}^4 f_y{}^4}{ \left| f_x{}^2 + i\frac{ f_x }{Q_x} (f-\alpha) - (f-\alpha)^2 \right|^2 \left| f_y{}^2 + i\frac{ f_y }{Q_y} \alpha - \alpha^2 \right|^2 } d\alpha \nonumber\\
&\approx \frac{8\pi^2}{T} \frac{\pi Q f_0{}^5 }{(f-|f_x-f_y|)^2} \left( \frac{1}{(f-2f_0)^2} + \frac{1}{(f+2f_0)^2} \right). \label{eq:GA}
\end{eqnarray}
Here $Q_x = Q_y \equiv Q \gg 1$ and $|f_x-f_y|\ll f_0$ is assumed for simplicity. 
From Eq. (\ref{eq:xP_power}) and (\ref{eq:GA}), the amplitude spectral density of the convolution $\left( \omega^2 \tilde{x} \right) \ast \tilde{\theta}_{\rm P} (f)$ is 
\begin{eqnarray}
\fl \hspace{20pt}
\sqrt{G_{\left( \omega^2 \tilde{x} \right) \ast \tilde{\theta}_{\rm P}}(f)} = \sqrt{\frac{8\pi^2}{T} \left< \left| \left( \omega^2 \tilde{x} \right) \ast \tilde{\theta}_{\rm P} (f) \right|^2 \right> }\nonumber\\
\fl \hspace{40pt}
= \frac{(2\pi \times 1\,{\rm Hz})^4}{g} G_{\rm seis}({1\,\rm Hz}) \sqrt{ \frac{\pi Q f_0{}^5 }{(f-|f_x-f_y|)^2} \left( \frac{1}{(f-2f_0)^2} + \frac{1}{(f+2f_0)^2} \right) }. \label{eq:G_xP}
\end{eqnarray}
This gives an approximation of nonlinear noise which is valid at low frequencies (below $\sim0.2$ Hz) as discussed in Sec. \ref{sec:reduction}.

\section*{References}
\bibliographystyle{unsrt}
\bibliography{TOBAbibl}

\end{document}